\documentclass[aip,pop,preprint,superscriptaddress,showpacs]{revtex4}
\usepackage{graphicx}
\usepackage{moreverb}
\usepackage{ifthen}
\usepackage{natbib}
\usepackage[linktocpage]{hyperref}
\usepackage{url,soul}

\newboolean{INTERNAL}
\setboolean{INTERNAL}{false}

\newcommand{\Caption}[2][{}]{%
\ifthenelse{\boolean{INTERNAL}}{\caption{#2 (#1)}}{\caption{#2}}}
\newcommand{\LSP}{\textsc{Lsp}}
\newcommand{\Alf}{{A}lfv\'{e}n}
\newcommand{\PROPACEOS}{P\textsc{ropaceos}}

\begin{document}

% Use the \preprint command to place your local institutional report
% number in the upper righthand corner of the title page in preprint mode.
% Multiple \preprint commands are allowed.
% Use the 'preprintnumbers' class option to override journal defaults
% to display numbers if necessary
%\preprint{}

%Title of paper
\title{Particle-in-cell simulations of collisionless shock formation via head-on merging of
two laboratory supersonic plasma jets}

% repeat the \author .. \affiliation  etc. as needed
% \email, \thanks, \homepage, \altaffiliation all apply to the current
% author. Explanatory text should go in the []'s, actual e-mail
% address or url should go in the {}'s for \email and \homepage.
% Please use the appropriate macro foreach each type of information

% \affiliation command applies to all authors since the last
% \affiliation command. The \affiliation command should follow the
% other information
% \affiliation can be followed by \email, \homepage, \thanks as well.
%\author{}
%\email[]{Your e-mail address}
%\homepage[]{Your web page}
%\thanks{}
%\altaffiliation{}
%\affiliation{}

\author{C.~Thoma}
\author{D.~R.~Welch}
\affiliation{Voss Scientific, LLC, Albuquerque, New Mexico 87108, USA}
\author{S.~C.~Hsu}
\affiliation{Los Alamos National Laboratory, Los Alamos, New Mexico 87545, USA}

\date{\today}

\begin{abstract}
We describe numerical simulations, using the particle-in-cell (PIC)
and hybrid-PIC code \LSP{} [T.~P.~Hughes et al., Phys.\ Rev.\ ST
  Accel.\ Beams~{\bf 2}, 110401 (1999)], of the head-on merging of two
laboratory supersonic plasma jets.  The goals of these experiments are
to form and study astrophysically relevant collisionless shocks in the
laboratory.  Using the plasma jet initial conditions (density $\sim
10^{14}$--$10^{16}$~cm$^{-3}$, temperature $\sim$ few eV, and
propagation speed $\sim 20$--100~km/s), large-scale simulations of jet
propagation demonstrate that interactions between the two jets are
essentially collisionless at the merge region. In highly resolved one-
and two-dimensional simulations, we show that collisionless shocks are
generated by the merging jets when immersed in applied magnetic fields
($B \sim 0.1$--1~kG)\@.  At expected plasma jet speeds of up to
100~km/s, our simulations do not give rise to unmagnetized
collisionless shocks, which require much higher velocities. The
orientation of the magnetic field and the axial and transverse density
gradients of the jets have a strong effect on the nature of the
interaction.  We compare some of our simulation results with those of
previously published PIC simulation studies of collisionless shock
formation.
\end{abstract}

% insert suggested PACS numbers in braces on next line
%
% insert suggested keywords - APS authors don't need to do this
%\keywords{}

%\maketitle must follow title, authors, abstract, \pacs, and \keywords
\maketitle

\section{Introduction}\label{sec:introduction}

Collisionless shocks \cite{sagdeev66,sagdeev91} play an important role
in energy transport and evolution of charged-particle distribution
functions in space and astrophysical environments.  Although
collisionless shocks in plasmas were first predicted in the 1950s
\cite{sagdeev66} and discovered in the 1960s \cite{ness64}, many
questions relating to the microscopic physics of collisionless shock
formation, evolution, and shock acceleration of particles to very high
energies remain unanswered \cite{plasma2010,wopa10}. Laboratory
studies of collisionless shocks have been conducted since the 1960s
\cite{podgornyi69,zakharov03}, but a recent renaissance of laboratory
collisionless shock experiments
\cite{woolsey01,horton07,ponomarenko08,romagnani08,constantin09,kuramitsu11,ross12,schaeffer12,swadling13}
stems from the fact that modern laboratory plasmas can satisfy key
physics criteria for the shocks to have ``cosmic relevance''
\cite{drake00}. Recently initiated experiments \cite{moser12} at Los
Alamos National Laboratory (LANL) aim to form and study
astrophysically relevant collisionless shocks via the head-on merging
of two supersonic plasma jets, each with order 10-cm spatial scale
size.  Compared to most other modern collisionless shock experiments
which use laser-produced or wire-array Z-pinch \cite{swadling13}
plasmas, the LANL experiment has larger shock spatial size (up to
30-cm wide and a few-cm thick) and longer shock time duration (order
10~$\mu$s) but somewhat lower sonic and Alfv\'en Mach numbers.  The
LANL experiment plans to have the capability to apply magnetic fields
of a few kG (via coils) that can be oriented either parallel or
perpendicular to the direction of shock propagation.  Obtaining
physical insights into and experimental data on collisionless shock
structure, evolution, and their effects on particle dynamics are the
primary reasons to conduct laboratory experiments on collisionless
shocks.  This paper reports results from particle-in-cell (PIC) and
hybrid-PIC numerical simulations, using the \LSP{} code
\cite{lspcode,hughes:99c}, that informed the design of the LANL
experiment and showed that collisionless shocks should appear with the
expected plasma jet parameters.

After a brief description of the LANL collisionless shock experiment,
the remainder of the paper describes single-jet propagation and one-
(1D) and two-dimensional (2D) PIC head-on merging jet simulations.
Our 1D magnetized simulations, in which the jets are immersed in an
applied magnetic field, are similar to those of Shimada and Hoshino
\cite{shimada:00} who performed 1D PIC simulations of magnetized shock
formation using a reduced ion-to-electron mass ratio and a reflection
boundary to model counter-propagating plasmas. We use the actual
hydrogen mass ratio and the actual hydrogen plasma parameters expected
in the LANL experiments, and we directly simulate both jets. This
gives us the flexibility to independently vary the properties ({\em
  e.g.}, the density profile) of the two jets without assuming any
symmetry. We have also performed 2D Cartesian merging simulations of
magnetized jets which allows us to consider the effects of the
orientation of the magnetic field and plasma density gradients with
respect to the jet propagation direction.  These simulations
demonstrate shock formation caused by the merging of magnetized jets
with Mach numbers as low as $\approx 1.5$, where the Mach number is
defined as \cite{tidman:71}
\begin{equation}
  M = \frac {v_1}{v_A \sqrt{2(1+5 \beta/6)}},
\label{eq:mach}
\end{equation}
where $v_1$ is the pre-shock jet velocity in the shock frame, $v_A =
B/\sqrt{2 \mu_o n_i m_i}$ is the \Alf{} velocity (in SI units) where
$B$ is the pre-shock magnetic field strength $B$, $n_i$ is the
pre-shock ion density, $m_i \gg m_e$ is the ion mass, and
\begin{equation}
  \beta = \frac {T_e+T_i}{m_i v_A^2},
\label{eq:beta}
\end{equation}
 where the $T_e$ and $T_i$ are the  pre-shock electron and ion
 temperatures in energy units.

In unmagnetized plasmas, collisionless shocks may also be formed by
the Weibel instability \cite{weibel59}. Simulations of this mechanism
were described by Kato and Takabe \cite{kato:08}, whose simulations
were also performed at a reduced mass ratio and were restricted to
relatively high velocities ($> 0.1c$). When using the hydrogen mass
ratio and a lower velocity ($\sim 100$ km/s as expected in the
experiment), we find no shock formation on relevant timescales (a few
$\mu$s).

The outline of the paper is as follows. In Sec.~\ref{sec:setup-model}
we describe the simulation setup and numerical models used. In
Sec.~\ref{sec:results}, we present \LSP\ simulation results of single
hydrogen jet propagation (Sec.~\ref{sec:single-h-jet} and
\ref{sec:fully-kinetic-single}) and 1D
(Sec.~\ref{sec:1d-magnetized-shock}) and 2D
(Sec.~\ref{sec:2d-simul-magn}) jet-merging with applied magnetic
fields.  Conclusions are given in Sec.~\ref{sec:conclusions}.

\section{Numerical models and simulation setup}\label{sec:setup-model}
\begin{figure}
\begin{center}
\includegraphics[clip=true, width=3truein]{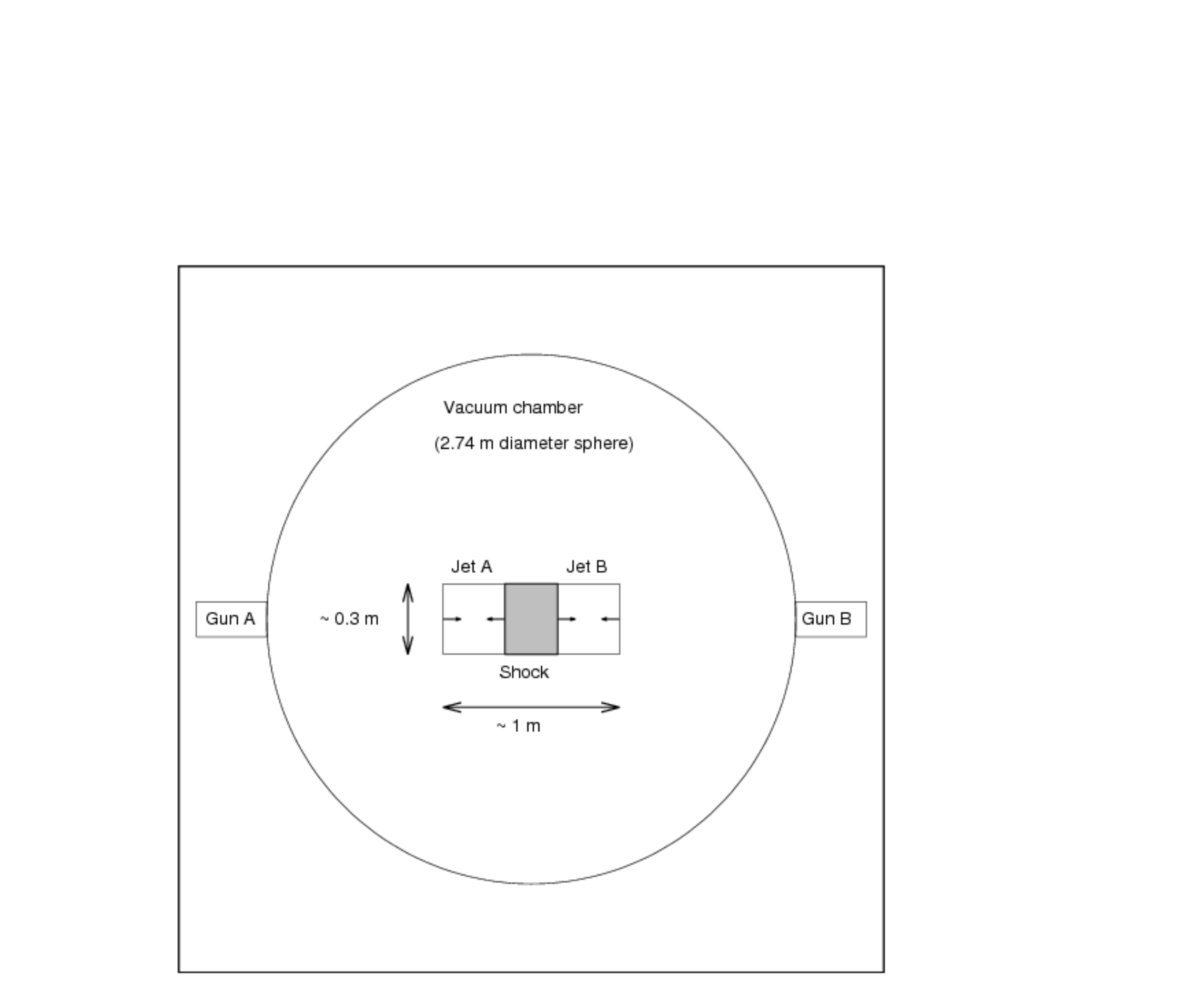}
\end{center}
\Caption[CT PC e:/reports/cless\_shock/exp.eps]{Schematic view of
  the experimental setup that is simulated in this paper.}
\label{fg:exp}
\end{figure}

The simulations described in this paper are based on the LANL
collisionless shock experiment \cite{moser12}, which uses
counter-propagating plasma jets formed and launched by plasma
railguns \cite{witherspoon11,thoma:11} mounted on opposite sides of a
2.74~m diameter spherical vacuum chamber (Fig.~\ref{fg:exp}).
Hydrogen, helium, and argon jets have been used in the experiments,
but we focus exclusively on hydrogen in this paper due to its ability to better
satisfy the physics criteria for cosmically relevant collisionless
shocks \cite{drake00}. Single-jet parameters and evolution have been
characterized experimentally \cite{hsu12pop} in preparation for using
an array of thirty such jets to form spherically imploding plasma
liners as a standoff compression driver for magneto-inertial
fusion \cite{hsu12ieee}. For these collisionless shock studies,
lower-density ($10^{14}$--$10^{16}$~cm$^{-3}$) and higher-velocity
(100~km/s) jets are desired; this is accomplished primarily by
reducing the injected mass for a given gun current.
%%%%%%%%%%%%%%%%%%%%%%%%%%%%%%%%%%%%%%%%%%%%%%%%%%%%%%%%%%%%%%%%%%%%%%%%
%  Schematic Figure of Experiment

%
%  E:\reports\cless_shock\exp.eps
%
% I got this figure from S. Hsu 
%  It's in the ppt file. But I redrew it for the paper
%
%  In this directory: gnuplot script .\plotpic
%
%  Gnuplot figures were huge.
%  Open eps file with GV, copy into paint crop and save as png file 
%  Very inelegant but it was the best way I found to do it.
%  LAST STEP: POP won't take png files. Use convert to go to pdf
%
%%%%%%%%%%%%%%%%%%%%%%%%%%%%%%%%%%%%%%%%%%%%%%%%%%%%%%%%%%%%%%%%%%%%%%%%

The approach used in this numerical study is two-fold. We initially
perform a large-scale simulation of a single jet propagating from the
end of the plasma gun to the center of the vacuum chamber. The
hydrogen jets emerge from the plasma gun with densities on the order
of $10^{14}$--$10^{16}$ cm$^{-3}$ and temperatures of a few eV. The
jets emerging from the guns will be few centimeters in size (on the
order of the railgun aperture) with masses of a few $\mu$g, and will
have a drift velocity $\sim 100$ km/s. But both must propagate on the
order of 1~m before merging can begin, during which time the density,
temperature, and equation-of-state (EOS) of the jet can change. The
single-jet propagation simulation models the time
evolution of the initial jet as it propagates through the
chamber. This 2D $r$--$z$ simulation must be run for several $\mu$s. This
requires using a fairly large timestep, for which $\omega_{pe} \Delta
t \gg 1$ at an initial plasma density $\sim 10^{16}$ cm$^{-3}$, where
$\omega_{pe}$ is the electron plasma frequency. Such simulations must
be done with a hybrid-PIC approach in which electron plasma
oscillations do need not to be resolved. 

However, a fully kinetic approach is required to model the formation
of shocks due to micro-instabilities induced by jet merging. As will
be seen in Sec.~\ref{sec:single-h-jet}, the hydrogen jets ejected from
the plasma guns drop considerably in density as they propagate to the
center of the chamber where the merging takes place. This density
reduction during propagation allows us to perform fully kinetic
explicit PIC merging simulations in 1D and 2D Cartesian coordinates,
in which electron timescales are resolved ($\omega_{pe} \Delta t <
1$). So these simulations, which are initialized with plasma
parameters obtained from the propagation simulation, are intended to
model only the merging process which occurs much later than the
ejection of the jets from the guns.

All of these simulations are performed using the hybrid-PIC code
\LSP{} \cite{hughes:99c}, which has been utilized widely and validated
for applications in many areas of beam and plasma physics, including
streaming instabilities \cite{genoni:04a,rose:07a} and Landau damping
\cite{rose:05a}. In addition to the traditional PIC paradigm, i.e., a
Maxwell-Vlasov solver for collisionless plasmas in which electron
length and time scales must be resolved, \LSP{} also contains
algorithms for dense plasma simulation and includes physics modules
for collisions (among charged and neutral species), EOS modeling, and
radiation transport. This flexibility available in the code makes it
useful for the two-fold simulation approach described above. All of
the simulation results presented later in the paper have been checked
to assure convergence of the relevant physics results with respect to
numerical parameters such as cell size, timestep, and particle number
per cell. To more fully focus on the physics results in the remaining
sections of the paper, we provide in this section details on the
models and numerical parameters used for the simulations.

\subsection{Setup of hybrid single jet propagation simulation}\label{sec:setup-single-jet}
The jet propagation simulation is performed in \LSP{} using a
quasi-neutral hybrid-PIC algorithm \cite{welch:11} which has fewer
constraints on the timestep. The ion macroparticles are kinetic. But
there are no electron macroparticles, as the ions carry fluid
information for the inertia-less electrons. The equation of motion for
the composite ion-electron macroparticle is given by
\begin{equation}
  m_i n_i \frac {D}{Dt} \vec{v} = \vec{J} \times \vec{B} -\nabla P_e
\label{eq:EOM}
\end{equation}
where $\vec{v}$ is the macroparticle velocity, $P_e$ is the electron
pressure, and $D/Dt$ is the full time derivative for the Lagrangian
macroparticle. The current is given by Ohm's law:
\begin{equation}
  \vec{J} = \sigma \left( \vec{E} + \vec{v}^* \times \vec{B} +\frac
      {\nabla P_e}{e n_e} \right)
\label{eq:ohm}
\end{equation} 
where $n_e$ is the electron density, $\sigma$ is the conductivity, and
$\vec{v}^*$ is the drift velocity gathered at the grid nodes. The
fields, current, densities, and electron pressure gradient are all
calculated at the nodes and then interpolated to the macroparticle
position when Eq.~(\ref{eq:EOM}) is applied. The full Maxwell's
equations are solved with the Ohm's law term included. Displacement
current is not dropped. Kinetic ions also undergo self-collisions
(ion-ion). It is for this reason that there is no ion pressure
contribution to Eq.~(\ref{eq:EOM}). Coulomb collisions between
electrons and ions species are included self-consistently through the
Spitzer conductivity in Eq.~(\ref{eq:ohm}). Particle energies are
advanced by the same method which is described in
Ref.~\cite{thoma:11}. The plasma EOS (plasma internal energy, charge
state, $\overline{Z} = n_e/n_i$, etc.)\ and opacity tables for
radiation transport are provided by the \PROPACEOS{} code
\cite{macfarlane:07}. Although \LSP{} includes a full radiation
transport algorithm \cite{thoma:11}, in this simulation we include
only photon emission and neglect absorption. This allows radiation to
be modeled as a simple energy sink on the fluid electron species. This
approximation is justified in the optically thin regime, which is well
satisfied for jets in the parameter regime under consideration.

The single jet propagation simulation is carried out in 2D $r$--$z$
cylindrical coordinates, as the jet is assumed to be azimuthally
symmetric. This allows for full spatial hydrodynamic expansion of the
jet in a 2D simulation. The simulation space is large enough to allow
for propagation of the initial jet from the exit of the gun to the
center of the vacuum chamber, and is bounded by perfectly conducting
metal walls. The cell size is $\Delta r = \Delta z = 0.5$~cm, and the
timestep is given by $c \Delta t = 32$~cm. The use of the
uniformly stable exact-implicit field solver allows the simulation to
be run with $c \Delta t \gg$ the cell size
\cite{zheng:00,zheng:01}. The initial plasma is characterized by $\sim
1000 $ ion macroparticles per cell. The results of the simulation are
discussed in Sec.~\ref{sec:single-h-jet}.

\subsection{Setup of fully kinetic 1D jet propagation and merging simulations}\label{sec:setup-explicit-1d}
The 1D Cartesian fully kinetic propagation and merging simulations are
initialized with input parameters based on results of the hybrid-PIC
single jet propagation simulation. As will be seen in
Sec.~\ref{sec:single-h-jet}, the jets at the chamber center have a
much lower density ($\sim 10^{13}$ cm$^{-3}$) than when ejected from
the plasma guns. This allows us do explicit kinetic PIC
simulations. We can also afford better spatial and temporal resolution
in 1D Cartesian coordinates: $\Delta x = 2 c \Delta t =
0.03$~cm. Since the timestep is small, an explicit field solution is
used rather than the exact-implicit solver.  Both electrons and ions
are modeled kinetically. Coulomb collisions are included for the
electron species, as are ion collisions, which are found to have a
negligible effect in this parameter regime.  The cell size and time
step given above resolve not only $\omega_{pe}$, but also the ion and
electron cyclotron frequencies ($\omega_{ci}$ and $\omega_{ce}$, respectively), and ion and electron skin depths
($d_i\equiv c/\omega_{pi}$ and $d_e\equiv c/\omega_{pe}$, respectively).  A
cloud-in-cell \cite{birdsall:91d} particle model allows $\Delta x >$
the Debye length. The 1D simulations described below are all run with
several thousand macroparticles per cell.  We assume fully stripped
hydrogen ions ($\overline{Z} = 1$) for simplicity. Some justification
for this assumption is given below. We also assume an ideal gas EOS
for the plasma jets and neglect radiation losses.

We initially perform a few fully kinetic single-jet propagation
simulations with and without magnetic fields to demonstrate the effect
of applied fields on single jet propagation. These results are
described in Sec.~\ref{sec:fully-kinetic-single}. In
Sec.~\ref{sec:1d-magnetized-shock} we discuss the results of 1D
fully kinetic simulations of two-jet merging with and without applied
magnetic fields. We consider the effects of varying magnetic field
strengths, as well as the effect of finite density gradients (with
scale length $L_d$), and the effect of the initial spatial separation,
or gap $L_g(0)$, between the two jets.

\subsection{Setup of fully kinetic 2D jet merging simulations}\label{sec:setup-explicit-2d}
The last group of simulations considered in this paper are in 2D
Cartesian coordinates. We simulate the 2D jets in the $x$--$y$~plane,
with the jets propagating in the $x$ direction. The total $y$ extent
is many $d_i$ wide, and periodic boundaries are imposed at
minimum and maximum values of $y$. To maintain reasonable runtimes, 2D
simulations performed at realistic length scales (jet lengths $L_j \sim
10$--100~cm) require coarser spatial resolution ($\Delta x = \Delta y
= 2 c\Delta t = 0.24$~cm) and a smaller number of particles per cell
(tens rather than hundreds) than were possible in 1D\@. Coulomb
collisions are included, but we again assume an ideal EOS and neglect
radiation transport (both photon emission and absorption).

In Sec.~\ref{sec:2d-simul-magn} we consider the results of 2D Cartesian simulations of
counter-propagating jets in perpendicular magnetic fields. As in 1D,
the jet propagation remains in the $x$ direction, and we simulate the
2D jets in the $x$--$y$~plane. The total $y$ extent is many $d_i$
wide, and periodic boundaries are imposed at minimum and
maximum values of $y$.

In these simulations we find some slow numerical heating of the
electrons at later times due to the coarse spatial resolution of the
grid ($\Delta x \sim c/\omega_{pe}$). However, if the simulation
duration does not exceed $\sim 1$~$\mu$s, the energy conservation
remains good to within a few percent. High fidelity simulations over
longer time scales will require better spatial resolution and larger
particle numbers. This will require the use of more processors than
were available.

\section{Simulation results}
\label{sec:results}

\subsection{Hybrid-PIC simulation of single hydrogen plasma jet propagation}
\label{sec:single-h-jet}

\begin{figure}
\begin{center}
\includegraphics[clip=true, width=12.0cm]{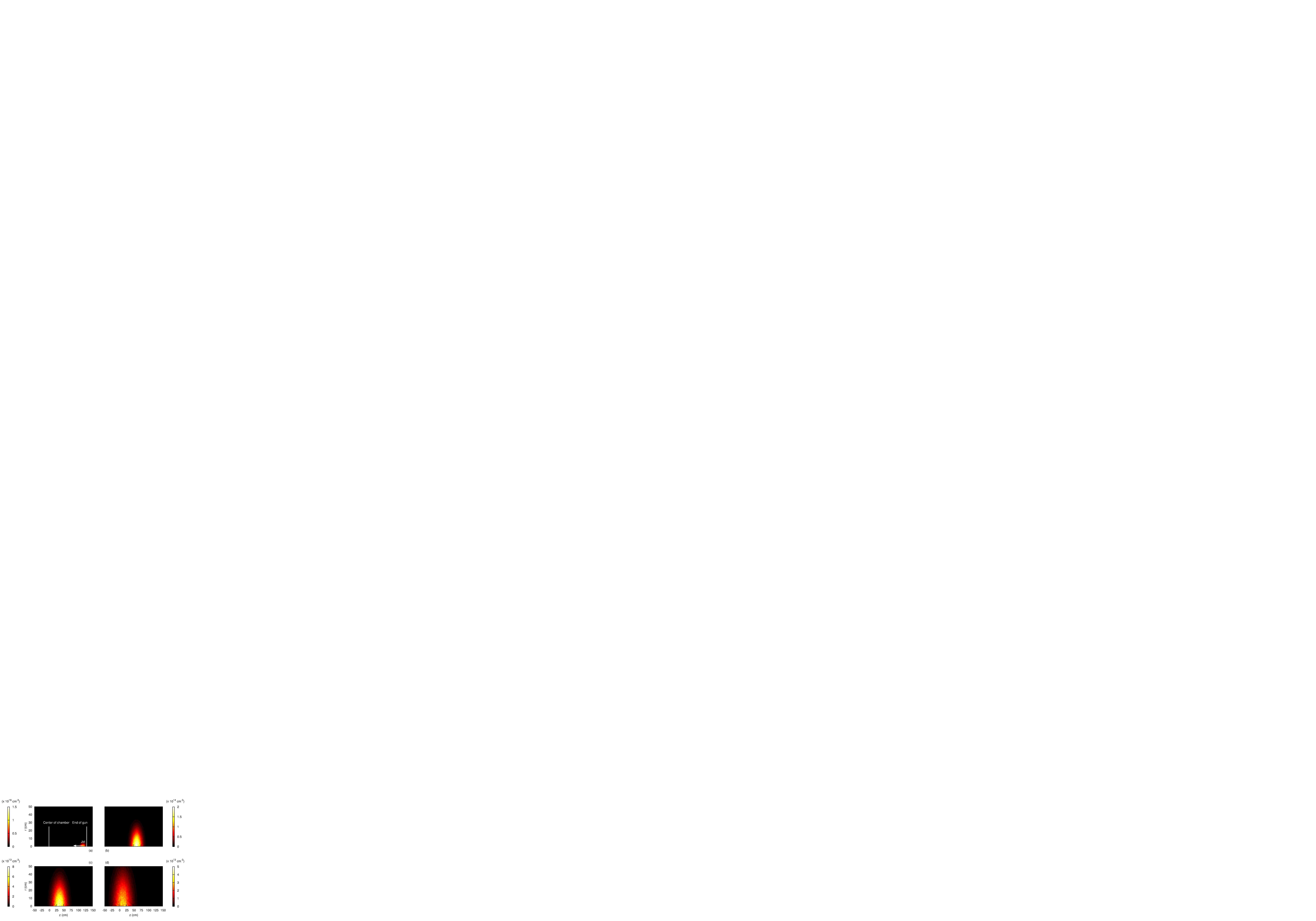}
\end{center}
\Caption[CT PC e:/reports/cless\_shock/ni\_trans.eps]{Snapshots of ion
  density in a 2D $r$--$z$ hybrid-PIC simulation of hydrogen jet propagation
  from the end of the plasma gun to the center of the
  chamber. Initial jet parameters are given in
  Table~\ref{tab_trans}. Densities are shown at $t =$ (a)~0, (b)~3800,
  (c)~5400, and (d)~7000~ns.}
\label{fg:ni_trans}
\end{figure}

Hybrid-PIC \LSP{} simulations were performed of a single hydrogen
plasma jet propagating from the railgun nozzle to the center of the
chamber in order to connect the plasma jet parameters at the railgun
exit with those in the region of head-on jet merging. Details on the
simulation setup and numerical methods are given in
Sec.~\ref{sec:setup-single-jet}. The initial ion density $n_i$ profile can
be in seen in the upper left plot in Fig.~\ref{fg:ni_trans}. At $t =
0$ the single jet is assumed to have a peak density $n_i(t=0) =
10^{16}$ cm$^{-3}$ (total mass $\sim 10$ $\mu$g), electron and ion temperatures $T_e(0)=T_i(0) =
5$~eV, and $v_j =150$~km/s (in the $-z$ direction). The initial jet
parameters are also given in Table~\ref{tab_trans}.

\begin{figure}
\begin{center}
\includegraphics[clip=true, width=12.0cm]{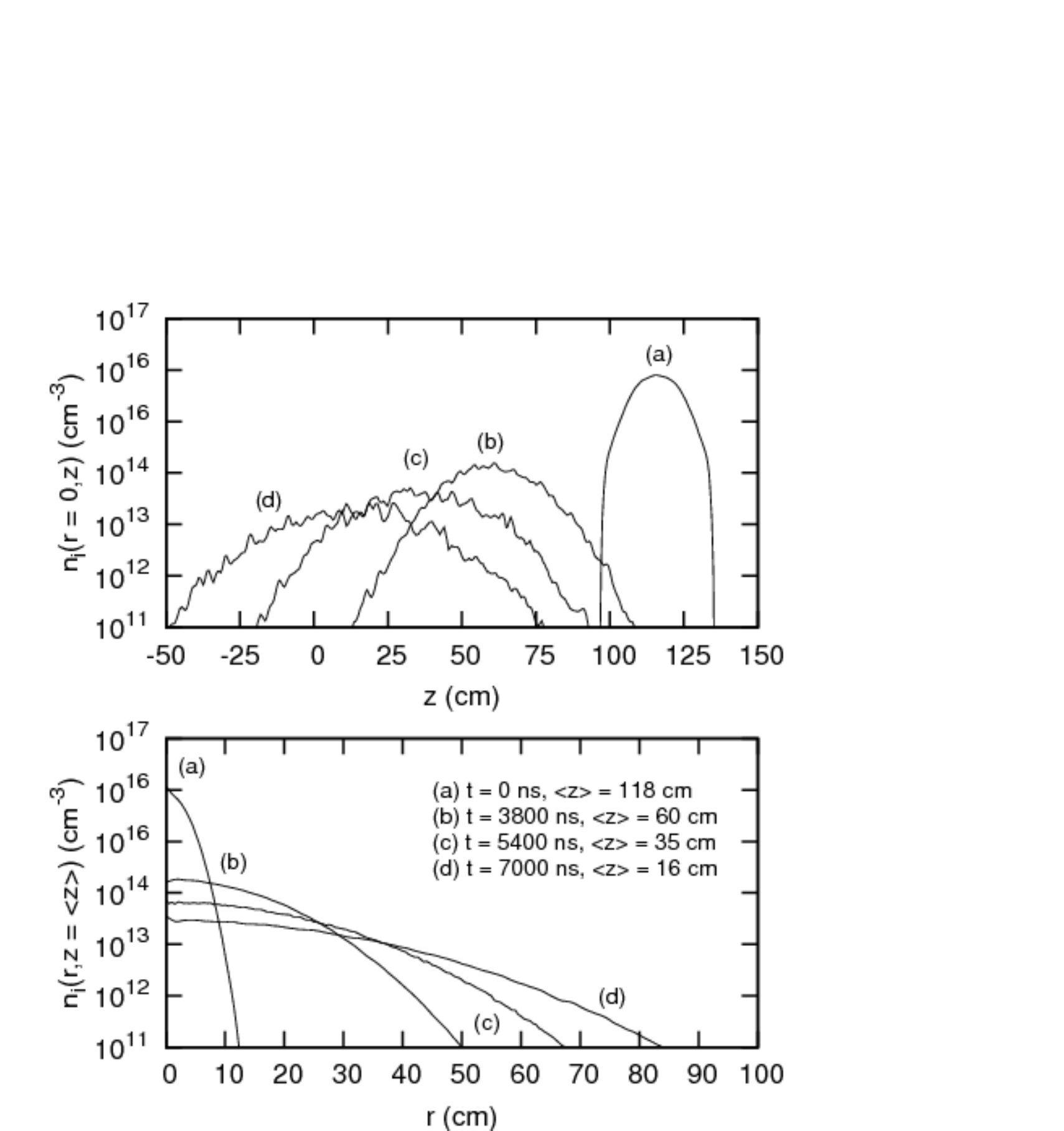}
\end{center}
\Caption[CT PC e:/reports/cless\_shock/n\_trans\_lo.eps]{Line-outs of
  ion density in a 2D $r$--$z$ hybrid-PIC simulation of hydrogen jet
  propagation from the end of the plasma gun to the center of the
  chamber. Densities are shown at $t =$ (a)~0, (b)~3800, (c)~5400,
  and (d)~7000~ns. (Top)~Line-outs vs.\ $z$ at $r
  =0$. (Bottom)~Line-outs vs.\ $r$ at $z = \langle z(t)\rangle$, which is the instantaneous, longitudinal jet center-of-mass.}
\label{fg:n_trans_lo}
\end{figure}

The time evolution of the jet propagation can be seen in
Fig.~\ref{fg:ni_trans}, which shows $n_i$ contours at $t = 0$, $3.8$,
$5.4$, and $7$~$\mu$s. Figure~\ref{fg:n_trans_lo} shows $n_i$
line-outs at the same times. The approximate plasma parameters of the
jet at the center of the chamber ($z = 0$~cm) are also given in
Table~\ref{tab_trans}. So this simulation determines the approximate
parameter regime of the individual jets after they have propagated to
the center of the chamber and begun to merge: $n_i \simeq
10^{13}$~cm$^{-3}$, $v_j \simeq 150$~km/s, $T_e \simeq T_i \simeq
1$~eV, $\overline{Z} = n_e/n_i \sim$ 1, and $L_j \sim 50$ cm. We note
that from spectroscopic data obtained from the LANL experiment the
plasma density at the chamber center can be inferred to be below
$10^{14}$ cm$^{-3}$, which is consistent with the simulation
result. The experimental jets are expected to be somewhat colder when
emerging from the guns than the 5-eV value used in the
simulations. But in the simulation results radiation cooling quickly
causes the jet temperature to drop. Nonetheless, we expect the amount
of density decay seen in the simulation to be an upper bound on the
experiments.

Based on the results of this simulation, we have chosen a set of
simplified plasma parameters to be used for the fully kinetic
simulations discussed in the following subsections. These values are
given in Table~\ref{tab_merge}.  Using these parameters we can
estimate the Coulomb collision frequency for the jets. To observe
collisionless merging of the jets, it is necessary that the inter-jet
ion collision time be much larger than the jet interaction time
$L_j/v_j$. For two counter-propagating ion beams ($v_j \gg v_{th,i}$,
the ion thermal velocity), the Spitzer collision frequency is
proportional to $|2 v_j|^{-3}$ \cite{rambo:95}. Using the parameters
above, we find $\nu_{ii}^{-1} \sim 10^{-4}$~s. The ion stopping time
due to collisions with electrons in the opposing jet is $\nu_{ie}^{-1}
\sim 10^{-5}$~s, while the jet interaction time $L_j/v_j \sim
10^{-6}$~s. So this simulation result demonstrates that these
counter-propagating jets will indeed be in the collisionless regime
when the jets merge at the center of the chamber, allowing the LANL
facility to be used for the investigation of collisionless shock
formation.

%%%%%%%%%%%%%%%%%%%%%%%%%%%%%%%%%%%%%%%%%%%%%%%%%%%%%%%%%%%%%%%%%%%%%%%%
%  Figure of 2D radial sim of jet from end of gun to chamber center 

%
%  E:\reports\cless_shock\ni_trans.eps
%
%  Figure taken from Sargas ~/hyperv/merge/run198/ (lsp run dir)
%    gnuplot script ./splot_ff_bw  (data files listed in there) 
%   eps file ./ni_trans.eps
%
%   Dump data from 2D contour plots in p4. Have to do some
%  post-processing to get gnuplot to read them (see notes in gnuplot
%  script)
%  run little shell script ~/bin/p4fix.p4 This is hard coded for 200 x
%  200 contour plot (Lsp default)  
%
%  Gnuplot figures were huge.
%  Open eps file with GV, copy into paint crop and save as png file 
%  Very inelegant but it was the best way I found to do it.
%  LAST STEP: POP won't take png files. Use convert to go to pdf
%
%%%%%%%%%%%%%%%%%%%%%%%%%%%%%%%%%%%%%%%%%%%%%%%%%%%%%%%%%%%%%%%%%%%%%%%%

%%%%%%%%%%%%%%%%%%%%%%%%%%%%%%%%%%%%%%%%%%%%%%%%%%%%%%%%%%%%%%%%%%%%%%%%
%  Figure of 2D radial sim of jet from end of gun to chamber center 
%   Lineouts rather than contour plots. Suggested by SH
%
%
%  E:\reports\cless_shock\n_trans_lo.eps
%
%  Figure taken from Sargas ~/hyperv/merge/run198/ (lsp run dir)
%    gnuplot script ./he4plot  (sic, data files listed in there) 
%   eps file ./n_trans_lo.eps
%
%  Gnuplot figures were huge.
%  Open eps file with GV, copy into paint crop and save as png file 
%  Very inelegant but it was the best way I found to do it.
%  LAST STEP: POP won't take png files. Use convert to go to pdf
%
%%%%%%%%%%%%%%%%%%%%%%%%%%%%%%%%%%%%%%%%%%%%%%%%%%%%%%%%%%%%%%%%%%%%%%%%

%%%%%%%%%%%%%%%%%%%%%%%%%%%%%%%%%%%%%%%%%%%%%%%%%%%%%%%%%%%%%%%%%%%%%%%%
% Table of initial jet parameters and value at center of chamber
%
%
\begin{table}
\caption{Initial and approximate parameters at the center of the
  chamber ($z = 0, t\simeq 8400$ ns) in 2D $r$--$z$ jet propagation
  simulation (see also Figs.~\ref{fg:ni_trans} and
  \ref{fg:n_trans_lo}).}
\begin{ruledtabular}
\begin{tabular}{ccc}
Quantity & Railgun Exit & Chamber Center \\ \hline 
$n_e$ (cm$^{-3}$) & $10^{16}$ & $9 \times 10^{12}$ \\
$n_i$ (cm$^{-3}$) & $10^{16}$ & $2 \times 10^{13}$ \\
$T_e$ (eV) & $5$ & $1.25$  \\
$T_i$ (eV) & $5$ & $0.4$  \\
$v_j$ (km/s) & $150$ & $150$ \\
\end{tabular}
\end{ruledtabular}
\label{tab_trans}
\end{table}
%
%%%%%%%%%%%%%%%%%%%%%%%%%%%%%%%%%%%%%%%%%%%%%%%%%%%%%%%%%%%%%%%%%%%%%%%%

%%%%%%%%%%%%%%%%%%%%%%%%%%%%%%%%%%%%%%%%%%%%%%%%%%%%%%%%%%%%%%%%%%%%%%%%
% Table of Canonical merging jet parameters used in Explicit PIC simulations
%
%
\begin{table}
\caption{Initial jet parameters used in 1D and 2D fully kinetic
  propagation and merging simulations. For simulations with density
  gradients, the density value refers to the peak
  value.  Magnetic field values refer to the unperturbed applied
  field strength}
\begin{ruledtabular}
\begin{tabular}{cc}
Quantity & Value  \\ \hline 
$n_i = n_e$ (cm$^{-3}$) & $10^{13}$ \\
$T_i = T_e$ (eV) & 1  \\
$v_j$ (km/s) & 150 \\
$L_j$ (cm) & $\sim 50$ \\
$L_d$ (cm) & varies: $\sim [0, L_j]$ \\
$B$ (G) & varies: $\sim [0,1500]$ \\
$ \nu_{ii}^{-1}$ (s) & $\sim 10^{-4}$ \\
$ \nu_{ie}^{-1}$ (s) & $\sim 10^{-5}$ \\
$L_j/v_j$ (s) & $\sim 10^{-6}$ \\
\end{tabular}
\end{ruledtabular}
\label{tab_merge}
\end{table}
%
%%%%%%%%%%%%%%%%%%%%%%%%%%%%%%%%%%%%%%%%%%%%%%%%%%%%%%%%%%%%%%%%%%%%%%%%

\subsection{Fully kinetic single-jet simulations in 1D}\label{sec:fully-kinetic-single}
\begin{figure}
\begin{center}
\includegraphics[clip=true, width = 10 cm]{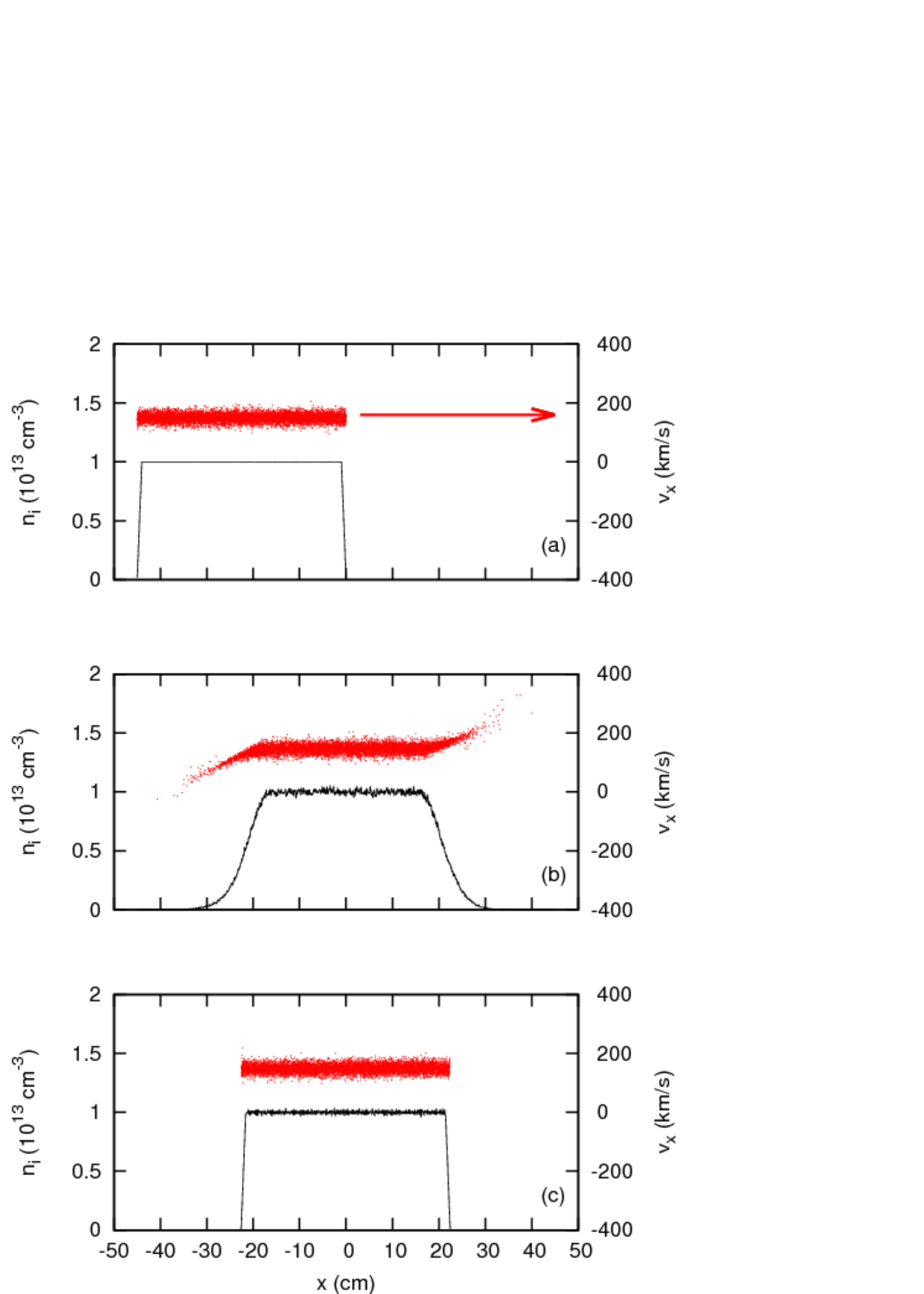}
\end{center}
\Caption[CT PC e:/reports/cless\_shock/n\_xpx.eps]{Fully kinetic, 1D,
  single-jet propagation simulation results for ion density $n_i$ and
  ion macroparticle $x$--$v_x$ phase-space data versus $x$ at (a)~$t =
  0$, and at $t = 1500$~ns with (b)~no applied magnetic field and
  (c)~$B_z = 1$~kG. }
\label{fg:n_xpx}
\end{figure}

We now consider fully kinetic, highly resolved 1D Cartesian simulations of jet propagation
in the merge region, with and without an applied perpendicular magnetic field. The
initial jet has a uniform density of $n = 10^{13}$~cm$^{-3}$. The
scale lengths of the initial gradients at the edges of the jet are on
the order of a cell width. This simplified density profile is used to
illustrate more clearly the effect of the magnetic field on single-jet
propagation. Such a flat-top profile also allows for very clear shock
structure to develop in the two-jet simulations discussed in
Sec.~\ref{sec:1d-magnetized-shock}. The full set of initial conditions
are given in Table~\ref{tab_merge}.  Although we imposed the initial
condition $\overline{Z} = 1$, there is no noticeable charge separation
between the kinetic electron and ion species as the simulation
advances, i.e., quasi-neutrality ($n_e \simeq n_i$) holds throughout.
The jet propagates in the $+x$ direction, with $x = 0$ (chamber center) representing
the merge point. Although there is no second jet in the
simulations in this subsection, we stress this choice of
origin now, as it is retained throughout the rest of the paper. The
initial $n_i$ profile and ion macroparticle $x$--$v_x$
phase-space distribution are shown in
Fig.~\ref{fg:n_xpx}(a).  Each red dot in the phase-space plots ($v_x$ is
plotted on the right axis) represents an ion macroparticle, each of
which has the same charge weight.

Figure~\ref{fg:n_xpx}(b) shows the $n_i$ profile and ion
velocity distribution at $t = 1500$~ns with no applied magnetic
field. The spreading of the $n_i$ profile in time at the jet edges
is due to ambipolar diffusion. Due to the initial density profile, the
peak density does not drop because sound waves have not had time to
propagate fully into the jet. The addition of an applied transverse
magnetic field of $1$~kG suppresses the spreading of the density
profile [Fig.~\ref{fg:n_xpx}(c)], explained as
follows. There is a motional electric field $E_y$ embedded in the bulk of the jet
which allows $\vec{E} \times \vec{B}$ drifting of the jet at
$\vec{v}_j$ through the magnetic field aligned in the $+z$
direction. But electrons which stray from the bulk are line-tied, restricting
ambipolar diffusion. The $E_y$, with the appropriate magnitude of $v_j B$, can be clearly
observed in the simulation results.

%%%%%%%%%%%%%%%%%%%%%%%%%%%%%%%%%%%%%%%%%%%%%%%%%%%%%%%%%%%%%%%%%%%%%%%%
%  Figure of 1D 1 Jet propagation with and without perpendicular B field
%
%  Figure taken from Sargas ~/hyperv/kinetic/weibel/run113 
%  run113 is lsp run dir for Bz = 0 case
%  ../run114 is lsp run dir for Bz = 1 kG case. 
%
%    gnuplot script ./he4plot  (data files listed in there) 
%   eps file ./n_xpx.eps
%
%
%  Gnuplot figures were huge.
%  Open eps file with GV, copy into paint crop and save as png file 
%  Very inelegant but it was the best way I found to do it.
%  LAST STEP: POP won't take png files. Use convert to go to pdf
%
%%%%%%%%%%%%%%%%%%%%%%%%%%%%%%%%%%%%%%%%%%%%%%%%%%%%%%%%%%%%%%%%%%%%%%%%

\subsection{Unmagnetized and magnetized jet merging in 1D}\label{sec:1d-magnetized-shock} 

We now consider merging of two counter-propagating jets to look for
evidence of shock generation, and two possible collisionless
shock mechanisms. The first is a
Weibel-mediated unmagnetized shock driven by a 2D
electromagnetic kinetic instability when the two jets merge
\cite{kato:08}. In 2D PIC simulations, Kato et al.
\cite{kato:08} found unmagnetized Weibel-mediated shocks for merging
jets with ``non-relativistic'' velocities $v_j = (0.1$--$0.45)c$.  Using
\LSP{} in similar simulations, we are able to reproduce the results of
Kato for collisionless merging jets with velocities in this velocity
range. But $0.1c$ is still 200 times faster than the 150~km/s jet velocities
expected in the LANL experiments. When scaling the
Weibel simulations down to the experimental parameter regime (and
using the actual hydrogen mass ratio $m_i/m_e = 1836$), we find no
evidence of shock formation or any strong jet interaction over $\mu$s
time scales, suggesting that the LANL experiment will not give rise
to unmagnetized collisionless shocks. The second possible mechanism is a magnetic shock which
occurs when jets merge while immersed in perpendicular magnetic fields
($\vec{B} \cdot \vec{v}_j = 0$) \cite{kato:08,tidman:71}. For the
remainder of the paper, we consider 1D and 2D simulation results for
merging jets in such applied magnetic fields.

The 1D merging simulations are initiated with the jets near the merge
point, after which they are allowed to collide. Our simulations are
similar to those of Shimada and Hoshino \cite{shimada:00}, but we
explicitly model both jets (rather than the reflecting particle
boundary used by Shimada {\em et al}.). This allows us to incorporate
arbitrary density profiles for each jet.  Again, we use the real value
of $m_i/m_e$ for the hydrogen jets rather than a reduced mass
ratio. Details on the simulation setup were given in
Sec.~\ref{sec:setup-explicit-1d}.

\subsubsection{Unmagnetized jet merging}\label{sec:unmagn-jet-merg}
\begin{figure}
\begin{center}
\includegraphics[clip=true, width=10.0cm]{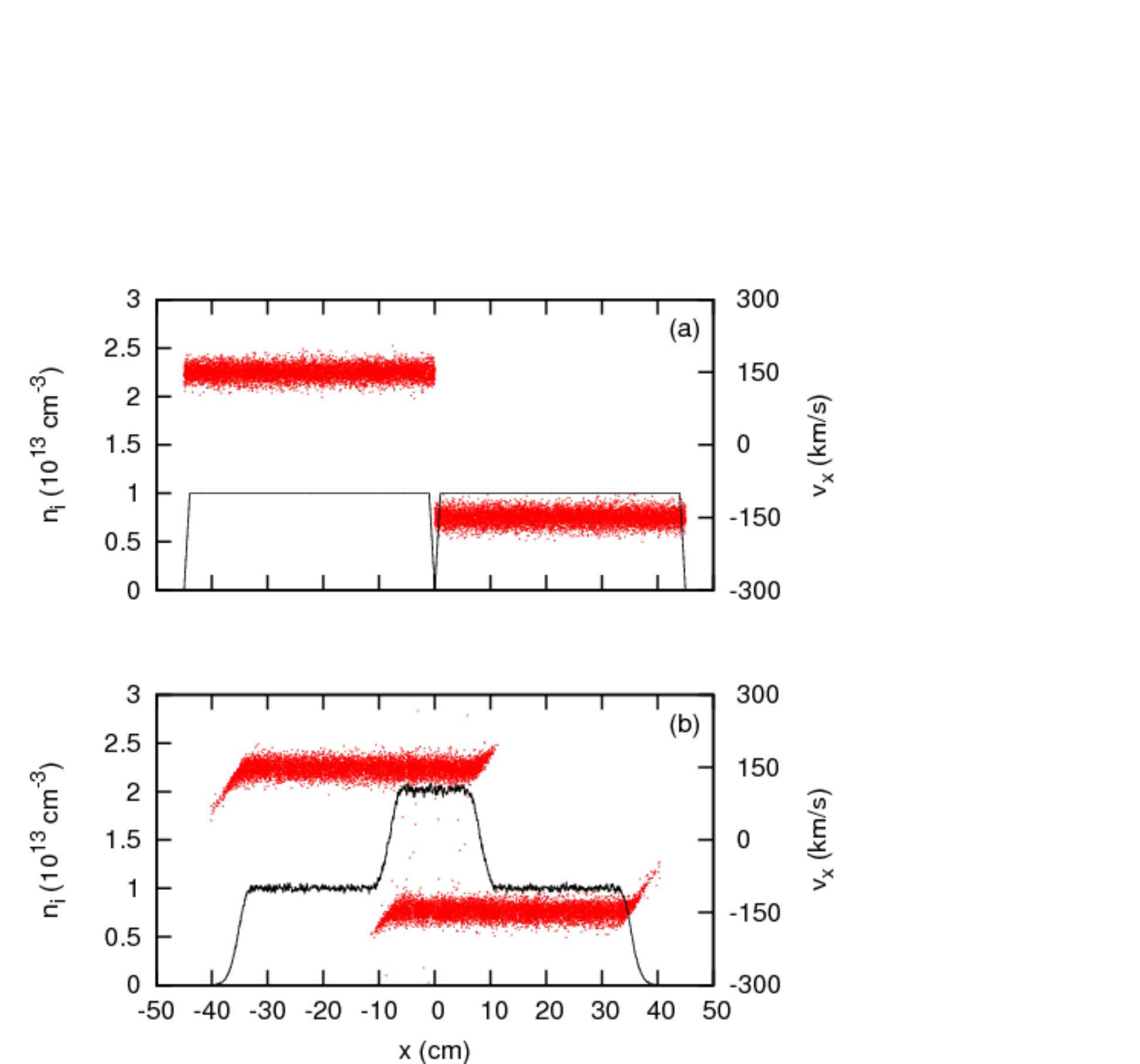}
\end{center}
\Caption[CT PC e:/reports/cless\_shock/n\_xpx\_2jet.eps]{Simulation
  (1D) results for two counter-propagating jets without an applied
  magnetic field. Initial jet parameters are given in
  Table~\ref{tab_merge}. Ion densities and particle $x$--$v_x$
  phase-space data are shown as a function of $x$ at (a)~$t = 0$ and
  (b)~$t = 1500$~ns.}
\label{fg:n_xpx_2jet}
\end{figure}

We now consider two counter-propagating jets in the absence of an
applied field. This situation in shown in Fig.~\ref{fg:n_xpx_2jet},
which shows $n_i$ and particle $x$--$v_x$ phase-space data as a
function of $x$ at various times. There is relatively little
interaction between jets in the absence of an applied magnetic
field. There are two distinct ion populations in the phase-space plot
representing each of the two jets, and the density doubles where the
two jets interpenetrate. This behavior demonstrates that ions are
effectively collisionless in this parameter regime (recall that
Coulomb collisions are fully included in the simulation and their
effect, if significant, would be visible in the ion phase-space). This
is consistent with the estimates given for the jet interaction time
and ion collision times in Table~\ref{tab_merge}. We also note that
since this is a 1D simulation, there is no possibility here of
observing the Weibel-driven shock, which requires at least 2D
modeling.

%%%%%%%%%%%%%%%%%%%%%%%%%%%%%%%%%%%%%%%%%%%%%%%%%%%%%%%%%%%%%%%%%%%%%%%%
%  Figure of 1D 2 counter-propagating jets no field
%
%  Figure taken from Sargas ~/hyperv/kinetic/weibel/run111
%  run111 is lsp run dir 
%
%    gnuplot script ./he4plot  (data files listed in there) 
%   eps file ./n_xpx_2jet.eps
%
%
%  Gnuplot figures were huge.
%  Open eps file with GV, copy into paint crop and save as png file 
%  Very inelegant but it was the best way I found to do it.
%  LAST STEP: POP won't take png files. Use convert to go to pdf
%
%%%%%%%%%%%%%%%%%%%%%%%%%%%%%%%%%%%%%%%%%%%%%%%%%%%%%%%%%%%%%%%%%%%%%%%%

\subsubsection{Magnetized jet merging}\label{sec:magn-jet-merg}
\begin{figure}
\begin{center}
\includegraphics[clip=false, width=15.0cm]{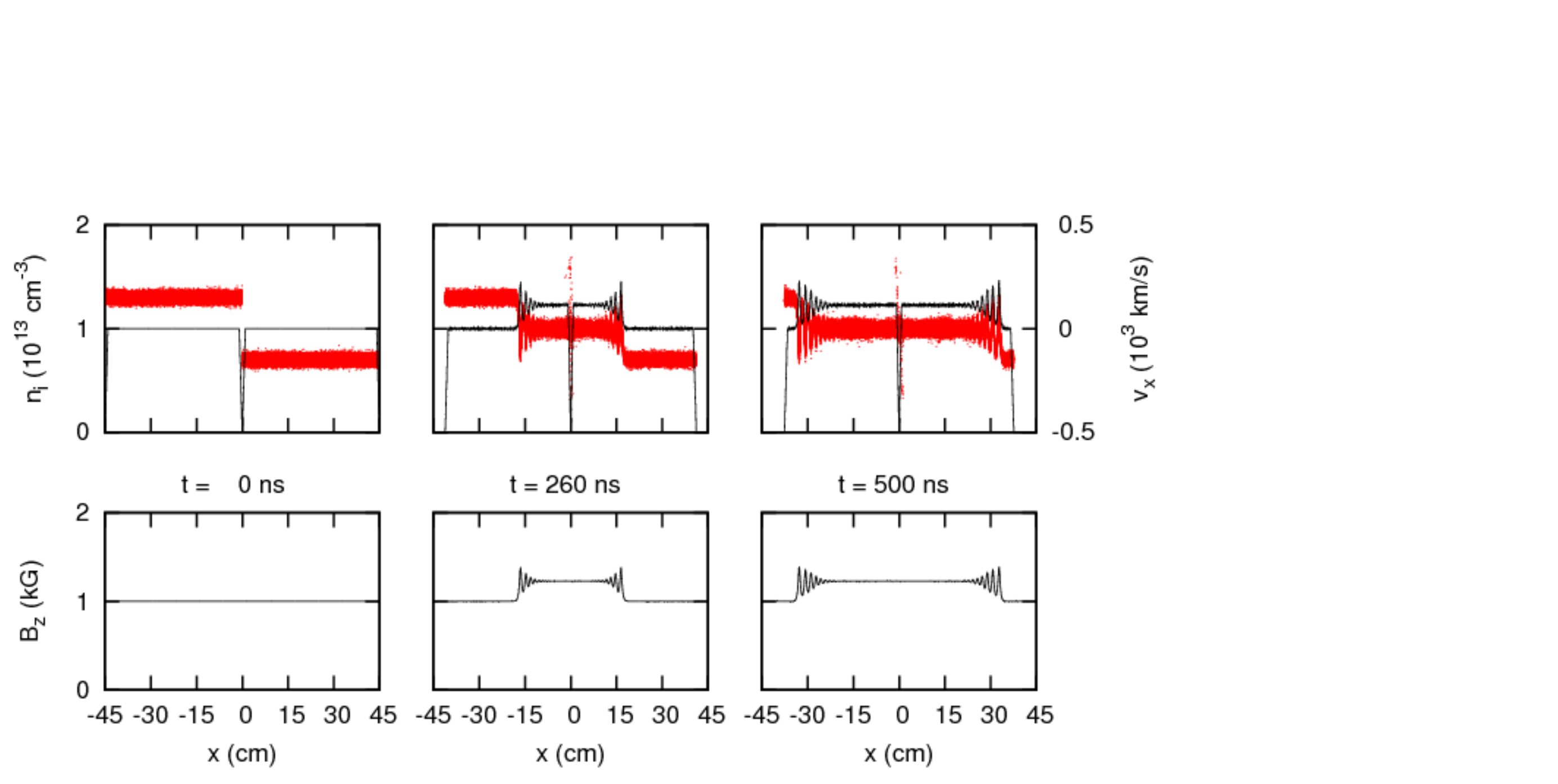}
\end{center}
\Caption[CT PC e:/reports/cless\_shock/n\_xpx\_2jet\_1kg.eps]{Simulation (1D)
results for two counter-propagating jets with an applied
  perpendicular field (in $+z$ direction) of $1$~kG\@. Initial jet
  parameters are given in Table~\ref{tab_merge}. Ion density, $B_z$,
  and ion macroparticle $x$--$v_x$ phase-space data are shown as a
  function of $x$ at (left column) $t = 0$, (middle column) $t = 260$~ns, and (right column)
  $t = 500$~ns.}
\label{fg:n_xpx_2jet_1kg}
\end{figure}

We now repeat the two jet simulation from
Sec.~\ref{sec:unmagn-jet-merg} but add an applied magnetic field, $B_z
= 1000$~G. Results from this simulation are shown in
Fig.~\ref{fg:n_xpx_2jet_1kg}. In this case there is clear propagation
of a shock wave. There is a finite width shock transition
region, with clear pre- and post-shocked regions with roughly constant
density and field values for each region. The persistence of the
density null at $x = 0$ can be explained as follows. Each jet travels
in a motional field $E_y \sim v_x B_z$, which has opposite signs for
the two jets ($v_x = \pm v_j$). At the merge point, $x =0$, the
motional fields cancel and the plasma can no longer propagate in the
magnetic field. Alternately, the enhanced magnetic field in the
shocked region is supported by plasma currents at the shock edge. The
$\vec{J} \times \vec{B}$ force decelerates the plasma from a drift
velocity of $v_j$ in the pre-shocked region to zero in the
post-shocked region. This can be seen in the phase-space plots. Since
the plasma is stationary behind the shock, the density profile remains
relatively constant in time. At early times, as the jets begin to
merge, the electric field $E_x$, which establishes the current flow,
can be seen in the simulation results. This field also causes the
acceleration of a small population of ions near $x = 0$~cm, which can
be seen in Fig.~\ref{fg:n_xpx_2jet_1kg}.

We now briefly discuss the electron behavior. We have already
mentioned that quasi-neutrality is maintained throughout all of the
simulations. There is no noticeable charge separation on length scales
greater than a cell width. But the electron velocity distribution does
not exhibit the clear structure which is observed in the ion
phase-space plots. There is only a sharp discontinuity in the electron
velocity spread (temperature) at the shock edge. Any more detailed
shock structure in the electron phase-space data is obscured by the
thermal spread of the electrons in the simulations. For this reason we
present only the ion phase-space data. We also note
that we did {\em not} observe  strong electron acceleration on
$\mu$s time scales, as seen by Shimada {\em et al.}\ in their
scaled simulations.

We have demonstrated the development of a propagating shock wave in
simulations with a perpendicular magnetic field. We find {\em no}
interaction between the jets if a {\em parallel} magnetic field is
applied.  There is also {\em no} interaction for small values of
perpendicular magnetic field, i.e., when $v_A \ll v_j$.

%%%%%%%%%%%%%%%%%%%%%%%%%%%%%%%%%%%%%%%%%%%%%%%%%%%%%%%%%%%%%%%%%%%%%%%%
%  Figure of 1D 2 counter-propagating jets WITH field

%
%
%  Figure taken from Sargas ~/hyperv/kinetic/weibel/run109
%  run109 is lsp run dir 
%
%    gnuplot script ./bxpxplot  (data files listed in there) 
%   eps file ./n_xpx_2jet_1kg.eps
%
%
%  Gnuplot figures were huge.
%  Open eps file with GV, copy into paint crop and save as png file 
%  Very inelegant but it was the best way I found to do it.
%  LAST STEP: POP won't take png files. Use convert to go to pdf
%
%%%%%%%%%%%%%%%%%%%%%%%%%%%%%%%%%%%%%%%%%%%%%%%%%%%%%%%%%%%%%%%%%%%%%%%%

\subsubsection{Effect of varying magnetic field magnitude}\label{sec:effect-vary-magn}
\begin{figure}
\begin{center}
\includegraphics[clip=true, width=15.0cm]{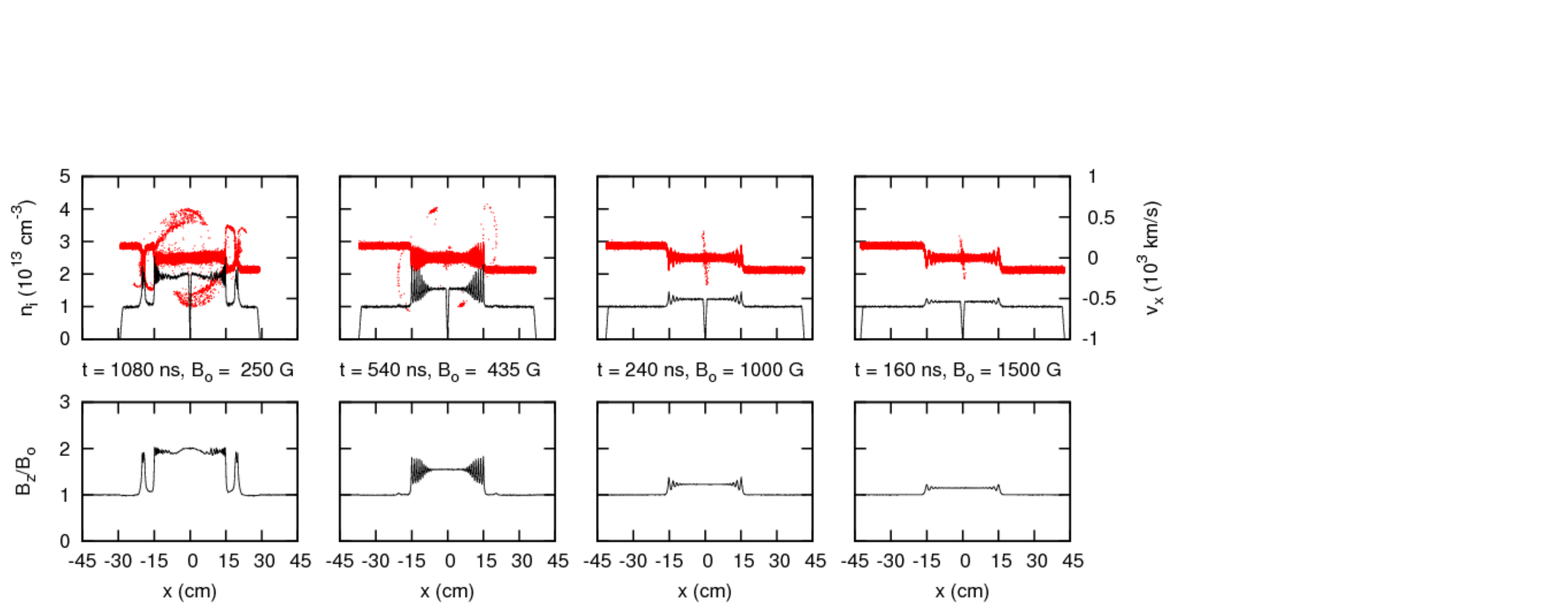}
\end{center}
\Caption[CT PC e:/reports/cless\_shock/n\_xpx\_2jet\_varyb.eps]{Simulation (1D)
results for two counter-propagating jets with varying
  applied perpendicular fields strengths (with magnitude $B_0$ in the
  $+z$ direction). Initial jet parameters are given in
  Table~\ref{tab_merge}. Ion density, $B_z$, and macroparticle $x$--$v_x$
  phase-space data are shown as a function of $x$. All plots are shown
  with the shock edge roughly at the same position, $|x| \sim 15$ cm.}
\label{fg:n_xpx_2jet_varyb}
\end{figure}

We now consider the effect of varying the field magnitude. In
Fig.~\ref{fg:n_xpx_2jet_varyb} we show results from a series of
simulations with varying values of applied field. The shock wave is
evident at all these field strengths, but other kinetic ion effects
are also visible.  The ripples in the density and field profiles in
the shock transition region ($|x| \sim [10,15]$ cm) have a spatial
period $\propto B$ (pre-shock field magnitude), and the ripple
amplitude grows with time. The shock transition region width is
approximately equal to $d_i$ and is roughly independent of $B$ and
$t$.  But simulations with longer jets are required to conclusively
show this scaling.  An explanation for the ion acceleration mechanism
at the initial merge point, $x \sim 0$, was explained
above. Individual ions accelerated at the shock edge, $|x| \sim
15$~cm, spin around in phase-space with a period $\tau \propto 1/B$
(but $\ll 2 \pi/ \omega_{ci}$) and a spatial extent of order the ion
Larmor radius $\sim r_{Li} \geq d_i$.

%%%%%%%%%%%%%%%%%%%%%%%%%%%%%%%%%%%%%%%%%%%%%%%%%%%%%%%%%%%%%%%%%%%%%%%%
%  Figure of 1D 2 counter-propagating jets with varying fields 2x4 array

%
%  Figure taken from Sargas ~/hyperv/kinetic/weibel/
%    gnuplot script ./bxpxplot  (data files listed in there) 
%
%  Run directories
%     ./run119 Bz = 250 G
%     ./run127 Bz = 435 G
%     ./run109 Bz = 1000 G
%     ./run120 Bz = 1500 G
%
%   eps file ./n_xpx_2jet_varyb.eps
%
%
%  Gnuplot figures were huge.
%  Open eps file with GV, copy into paint crop and save as png file 
%  Very inelegant but it was the best way I found to do it.
%
%%%%%%%%%%%%%%%%%%%%%%%%%%%%%%%%%%%%%%%%%%%%%%%%%%%%%%%%%%%%%%%%%%%%%%%%

\begin{figure}
\begin{center}
\includegraphics[clip=true, width=15.0cm]{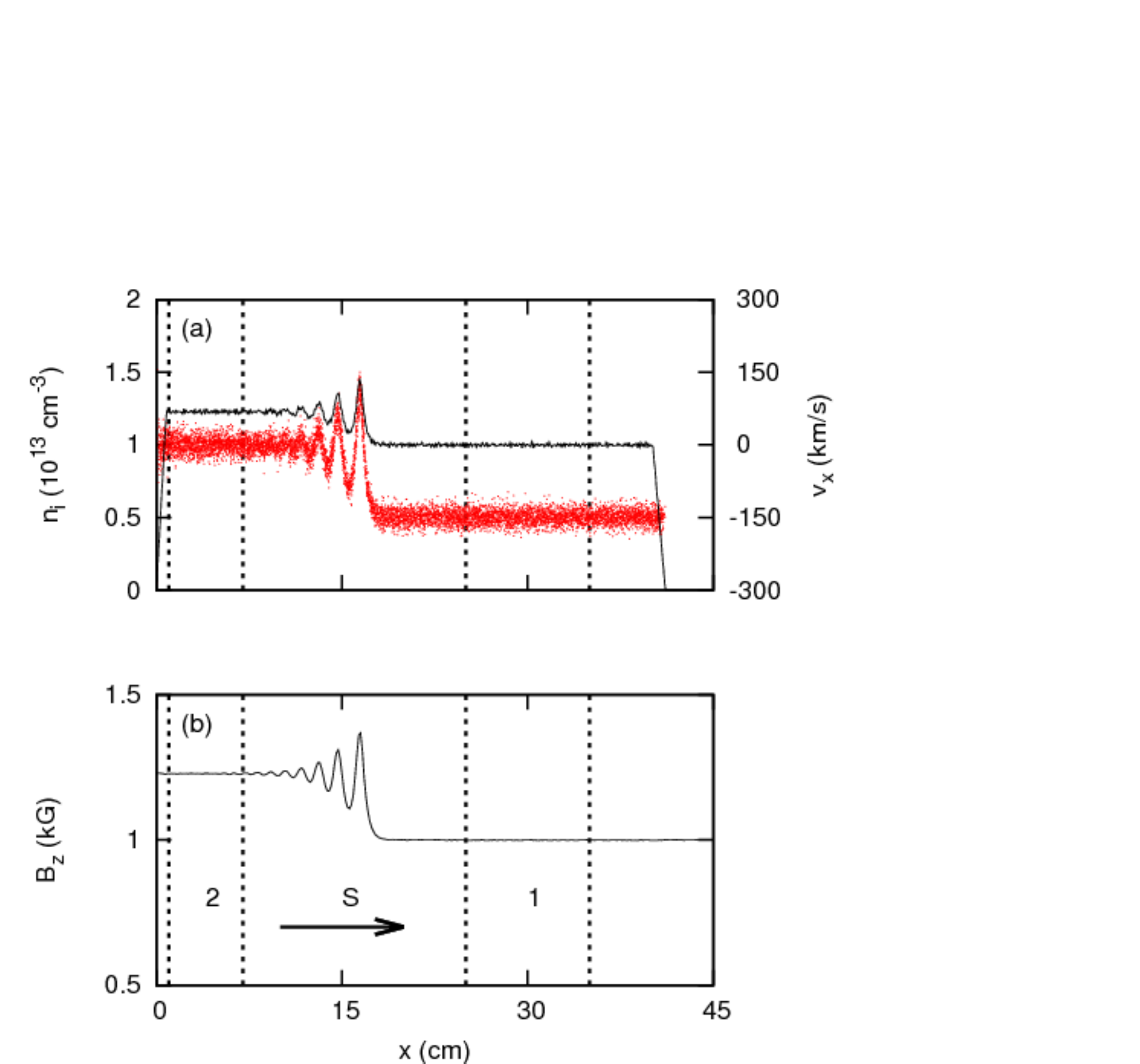}
\end{center}
\Caption[CT PC e:/reports/cless\_shock/shock\_zones.eps]{Simulation (1D)
  results for two counter-propagating jets with a $1$-kG perpendicular
  field in the $+z$ direction. Only the initially
  right-propagating jet is shown here. In the full simulation there is
  essentially mirror symmetry about $x = 0$. Initial jet
  parameters are given in Table~\ref{tab_merge}. (a)~Ion density and particle
  $x$--$v_x$ phase-space data, and (b)~$B_z$ are shown as a function
  of $x$ at $t = 260$~ns, a time well after the establishment of the
  propagating shock.}
\label{fg:shock_zones}
\end{figure}

For these simple 1D simulations, it is possible to compare our results
to the theoretical Rankine-Hugoniot conditions for collisionless
shocks in perpendicular fields. First, we define the regions of the
simulation space in Fig.~\ref{fg:shock_zones}. Region 1 is the
pre-shocked unperturbed region, and region 2 is the post-shocked
region, and we denote the transition region as ``S.'' In the
simulation frame, the S region propagates to the right in
Fig.~\ref{fg:shock_zones} at a shock velocity denoted as $v_s$. The
theoretical treatment given by Tidman and Krall \cite{tidman:71},
however, considers the shock frame, i.e., the frame where the
shock transition region is stationary. For a perpendicular magnetized
shock, $\vec{B}$ is perpendicular to both $\vec{v_1}$ and $\vec{v_2}$,
which are the pre- and post-shocked velocities (both in the $x$ direction), respectively, in the shock frame.
Assuming Maxwellian distributions for
ions and electrons, the Rankine-Hugoniot relations \cite{tidman:71}
are given by
\begin{equation}
  \frac {n_1}{n_2} = \frac {B_1}{B_2} = \frac {v_2}{v_1} = \frac
        {1}{8} \left( 1 + \frac {5 B_1^2 (1 + \beta)}{2 \mu_o m_i n_1 v_1^2}
                 + \left[
                   1 + \frac {5 B_1^2 (1 + \beta)}{2 \mu_o m_i n_i
                     v_1^2} + \frac {8 B_1^2}{\mu_o m_i n_1 v_1^2}
                   \right]^{1/2}
                \right),
\label{eq:HR}
\end{equation}
and the Mach number of the shock is given by Eq.~(\ref{eq:mach}).  We
note that there is a typographical error in the Rankine-Hugoniot
conditions in Ref.~\cite{tidman:71}, but in Eq.~(\ref{eq:HR}) the
error has been corrected. Also, since the theory assumes thermal
distributions for both electrons and ions, it does not account for any
of the ion kinetic effects seen in Fig.~\ref{fg:n_xpx_2jet_varyb}. We
note from Eq.~(\ref{eq:mach}) that in the limiting case where $\beta
\ll 1$, $M$ is the \Alf{} Mach number, $v_1/\sqrt{2} v_{A}$. In the
opposite limit ($\beta \gg 1$), $M$ becomes the sonic Mach number.

From the simulation results, the velocity ratio in the shock frame is
calculated by
\begin{equation}
  v_1/v_2 = (v_j+v_s)/v_s.
\end{equation}
The results for the density, magnetic field and velocity ratios, as
well as the Mach number, are shown Table~\ref{tab_hugo}. We see that
for all our merging simulations, we are in the low-$\beta$ regime. In
the 1D simulations, we always find $n_1/n_2 \simeq B_2/B_1 \simeq
v_1/v_2$.  There is good agreement with the theoretical values from
the Rankine-Hugoniot conditions, except at lower $B_z$ values which
are strongly perturbed by ion acceleration from the shock edge [a
kinetic effect not accounted for in Eq.~(\ref{eq:HR})]. This gives us
confidence that we can quantitatively describe shock phenomena using
PIC methods.

%%%%%%%%%%%%%%%%%%%%%%%%%%%%%%%%%%%%%%%%%%%%%%%%%%%%%%%%%%%%%%%%%%%%%%%%
%  Figure of 1D 2 counter-propagating jets. Show regions of shock
%
%
%  Figure taken from Sargas ~/hyperv/kinetic/weibel/run109
%    gnuplot script ./plotshock  (data files listed in there) 
%
%  Run directories
%     ./ Bz = 1000 G
%
%   eps file ./shock_zones.eps
%
%
%  Gnuplot figures were huge.
%  Open eps file with GV, copy into paint crop and save as png file 
%  Very inelegant but it was the best way I found to do it.
%  LAST STEP: POP won't take png files. Use convert to go to pdf
%
%%%%%%%%%%%%%%%%%%%%%%%%%%%%%%%%%%%%%%%%%%%%%%%%%%%%%%%%%%%%%%%%%%%%%%%%

%%%%%%%%%%%%%%%%%%%%%%%%%%%%%%%%%%%%%%%%%%%%%%%%%%%%%%%%%%%%%%%%%%%%%%%%
% Table of sims compared to Hugoniot
%
%
\begin{table}
\caption{Comparison of 1D simulation results to the Rankine-Hugoniot
  conditions of Eq.~(\ref{eq:HR}). Initial jet parameters are given in
  Table~\ref{tab_merge}}.
\begin{ruledtabular}
\begin{tabular}{cccccc}
$B_1 (G)$ & $\beta$ & $n_2/n_1 = B_2/B_1$ & $v_1/v_2$ & Theory & $M$ \\ \hline 
$250$ & $0.026$ & 2.0 & 2.09 & 1.79 & 1.65 \\
$350$ & $0.013$ & 1.7 & 1.69 & 1.63 & 1.51 \\
$435$ & $0.0085$ & 1.55 & 1.52 & 1.55 & 1.45 \\
$500$ & $0.0064$ & 1.48 & 1.44 & 1.48 & 1.41 \\
$1000$ & $0.0013$ & 1.22 & 1.22 & 1.26 & 1.22 \\
$1500$ & $0.0008$ & 1.15 & 1.15 & 1.19 & 1.15 \\
\end{tabular}
\end{ruledtabular}
\label{tab_hugo}
\end{table}
%
%%%%%%%%%%%%%%%%%%%%%%%%%%%%%%%%%%%%%%%%%%%%%%%%%%%%%%%%%%%%%%%%%%%%%%%%
\subsubsection{Effects of realistic density gradient at and gap size between jet leading edges}\label{sec:effects-real-dens}

\begin{figure}
\begin{center}
\includegraphics[clip=true, width=15.0cm]{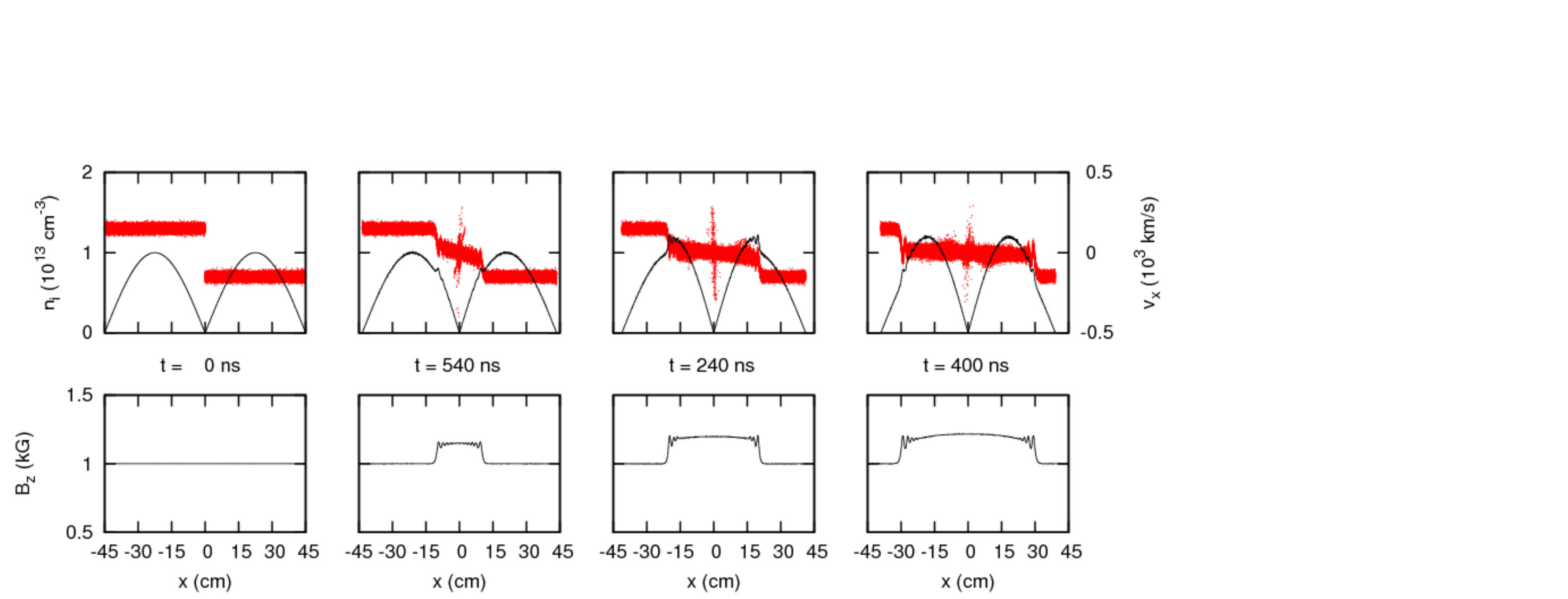}
\end{center}
\Caption[CT PC e:/reports/cless\_shock/n\_xpx\_2jet\_varyn.eps]{Simulation (1D)
results for two counter-propagating jets with finite
  density gradients in an applied perpendicular magnetic field (1 kG
  in the $+z$ direction). Initial jet parameters are given in
  Table~\ref{tab_merge}. Ion density, $B_z$, and particle $x$--$v_x$
  phase-space data are shown as a function of $x$.}
\label{fg:n_xpx_2jet_varyn}
\end{figure}

The previous 1D shock simulations were performed with very steep
density gradients (on the order of one or two cell widths), and a
uniform density in the bulk. These profiles were chosen because they
exhibited very clear shock structure and allowed direct comparison with 
the 1D theory. We now add more realistic density gradients. In the
experiment we expect density profiles such as those seen at later
times in Fig.~\ref{fg:ni_trans}, with $L_d\sim L_j$. The
applied field is $B_z = 1000$~G, and the maximum initial
density remains $10^{13}$ cm$^{-3}$. The initial density profile can
be seen at the top left of Fig.~\ref{fg:n_xpx_2jet_varyn}. The time
evolution of the density, particle phase-space, and magnetic field are
shown in the figure as well. We note that there is still a clear shock
wave structure visible when finite density gradients are included. Due
to the presence of initial density gradients, there are no constant
post-shock values of $n_2$ and $B_2$, but the sharp shock transition
region is evident.

%%%%%%%%%%%%%%%%%%%%%%%%%%%%%%%%%%%%%%%%%%%%%%%%%%%%%%%%%%%%%%%%%%%%%%%%
%  Figure of 1D 2 counter-propagating jets with density gradients 2x4 array

%
%  Figure taken from Sargas ~/hyperv/kinetic/weibel/run110
%    gnuplot script ./bxpxplot  (data files listed in there) 
%
%  Run directories
%     ./ Bz = 1000 G
%     ./run127 Bz = 435 G
%     ./run109 Bz = 1000 G
%     ./run120 Bz = 1500 G
%
%   eps file ./n_xpx_2jet_varyn.eps
%
%
%  Gnuplot figures were huge.
%  Open eps file with GV, copy into paint crop and save as png file 
%  Very inelegant but it was the best way I found to do it.
%  LAST STEP: POP won't take png files. Use convert to go to pdf
%
%%%%%%%%%%%%%%%%%%%%%%%%%%%%%%%%%%%%%%%%%%%%%%%%%%%%%%%%%%%%%%%%%%%%%%%%

\begin{figure}
\begin{center}
\includegraphics[clip=true, width=15.0cm]{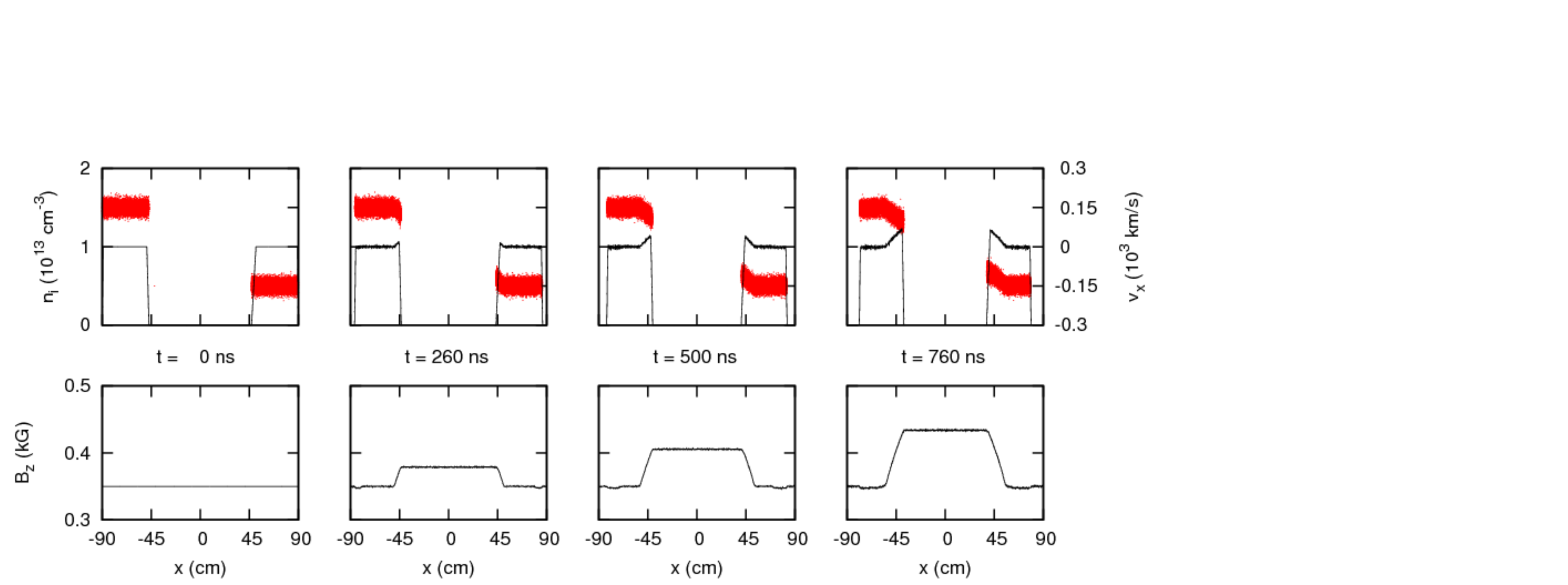}
\end{center}
\Caption[CT PC e:/reports/cless\_shock/n\_xpx\_2jet\_\_350sep.eps]{Simulation (1D)
results for two counter-propagating jets with a $350$~G
  perpendicular field in the $+z$ direction and an initial jet
  separation of $\sim 90$~cm.  (Top)~Ion density and particle $x$--$v_x$ phase-space
  data, and (bottom)~$B_z$ are shown as a function of $x$ at $t
  = 0$, 260, 500 and 760~ns.}
\label{fg:n_xpx_2jet_350sep.eps}
\end{figure}

All the 1D shock simulations discussed above were initialized with
two counter-propagating jets adjacent to each other (see the top left
of Fig.~\ref{fg:n_xpx_2jet_1kg}). This was done essentially to
minimize the run-time required to observe the interaction. We now wish
to consider the effect of an initial jet separation. We repeat the
simulation with constant bulk density and sharp density gradients and
$B_z = 350$ G (with shock parameters given in Table~\ref{tab_hugo}),
but now introduce a large initial jet separation $L_g(0) \sim L_j$. The initial geometry is shown in
Fig.~\ref{fg:n_xpx_2jet_350sep.eps}, as well as the time evolution of
ion density, ion particle phase-space, and magnetic field.
The interaction seen in this figure is not a shock wave.  At these
early times there is no clear post-shocked region:  $n_2$ and $B_2$
never reach constant steady-state values (despite the uniform initial
density). Moreover, the disturbance propagates into the jet at a speed
$\sim v_A$ (in the jet rest frame).

%%%%%%%%%%%%%%%%%%%%%%%%%%%%%%%%%%%%%%%%%%%%%%%%%%%%%%%%%%%%%%%%%%%%%%%%
%  Figure of 1D 2 counter-propagating jets. 350 G, 90 cm initial separation
%
%
%  Figure taken from Sargas ~/hyperv/kinetic/weibel/run132
%    gnuplot script ./bxpxplot  (data files listed in there) 
%
%  Run directories
%     ./ Bz = 350 G 90 cm separation
%
%   eps file ./n_xpx_2jet_350sep.eps
%
%
%  Gnuplot figures were huge.
%  Open eps file with GV, copy into paint crop and save as png file 
%  Very inelegant but it was the best way I found to do it.
%
%%%%%%%%%%%%%%%%%%%%%%%%%%%%%%%%%%%%%%%%%%%%%%%%%%%%%%%%%%%%%%%%%%%%%%%%

\begin{figure}
\begin{center}
\includegraphics[clip=true, width=15.0cm]{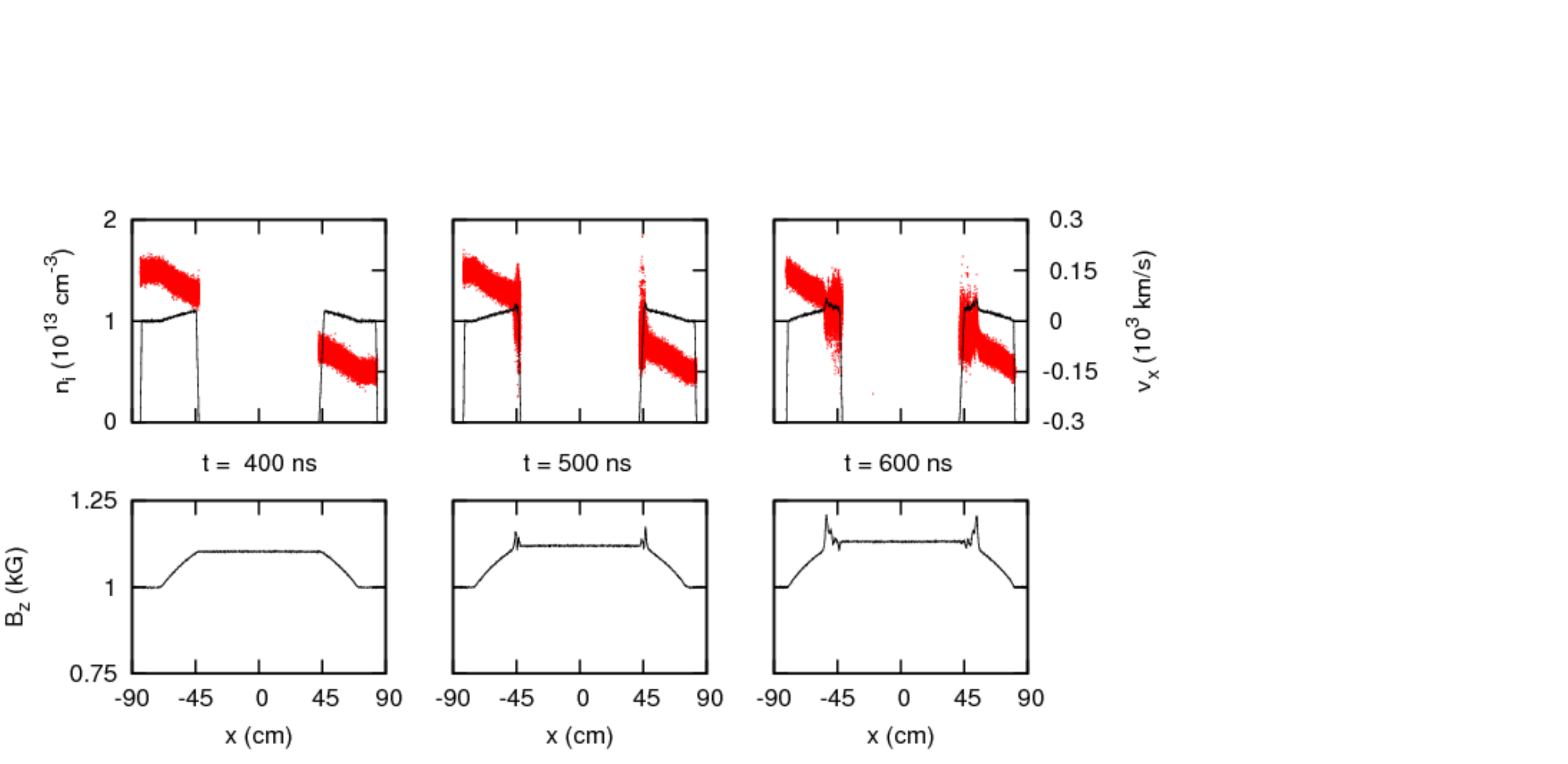}
\end{center}
\Caption[CT PC e:/reports/cless\_shock/n\_xpx\_2jet\_\_1000sep.eps]{Simulation (1D)
results for two counter-propagating jets with a $1$~kG
  perpendicular field in the $+z$ direction and an initial jet
  separation of $90$~cm.  Ion density, particle $x-v_x$ phase-space
  data (top), and $B_z$ (bottom) are shown as a function of $x$ at $t
  = 400$, $500$, and $600$~ns.}
\label{fg:n_xpx_2jet_1000sep.eps}
\end{figure}

We now also repeat the simulation at 1000 G with a larger initial jet
separation. Some results are shown in
Fig.~\ref{fg:n_xpx_2jet_1000sep.eps}. At early times we again see the
Alfv\'{e}nic disturbance propagate into the jets.  But later, at about
500 ns in this case, a shock wave begins to form, as evidenced by the
sharp discontinuity in the magnetic field and the broad spread in ion
velocities. Upstream of the shock, we notice deceleration of the
ions in the \Alf{}ic propagation region. In a series of such
simulations, we have noticed that the shock wave is initiated only when
there is a significant population of ions with $v_x \sim 0$. The shock
wave was not observed in the $350$~G simulation
(Fig.~\ref{fg:n_xpx_2jet_350sep.eps}) because it was not run long
enough for any ions to be slowed to small enough values of $v_x$.

%%%%%%%%%%%%%%%%%%%%%%%%%%%%%%%%%%%%%%%%%%%%%%%%%%%%%%%%%%%%%%%%%%%%%%%%
%  Figure of 1D 2 counter-propagating jets. 1000 G, 90 cm initial separation
%
%
%  Figure taken from Sargas ~/hyperv/kinetic/weibel/run131
%    gnuplot script ./bxpxplot  (data files listed in there) 
%
%  Run directories
%     ./ Bz = 1000 G 90 cm separation
%
%   eps file ./n_xpx_2jet_1000sep.eps
%
%
%  Gnuplot figures were huge.
%  Open eps file with GV, copy into paint crop and save as png file 
%  Very inelegant but it was the best way I found to do it.
%  LAST STEP: POP won't take png files. Use convert to go to pdf
%
%%%%%%%%%%%%%%%%%%%%%%%%%%%%%%%%%%%%%%%%%%%%%%%%%%%%%%%%%%%%%%%%%%%%%%%%

\begin{figure}
\begin{center}
\includegraphics[clip = true, width = 10 cm]{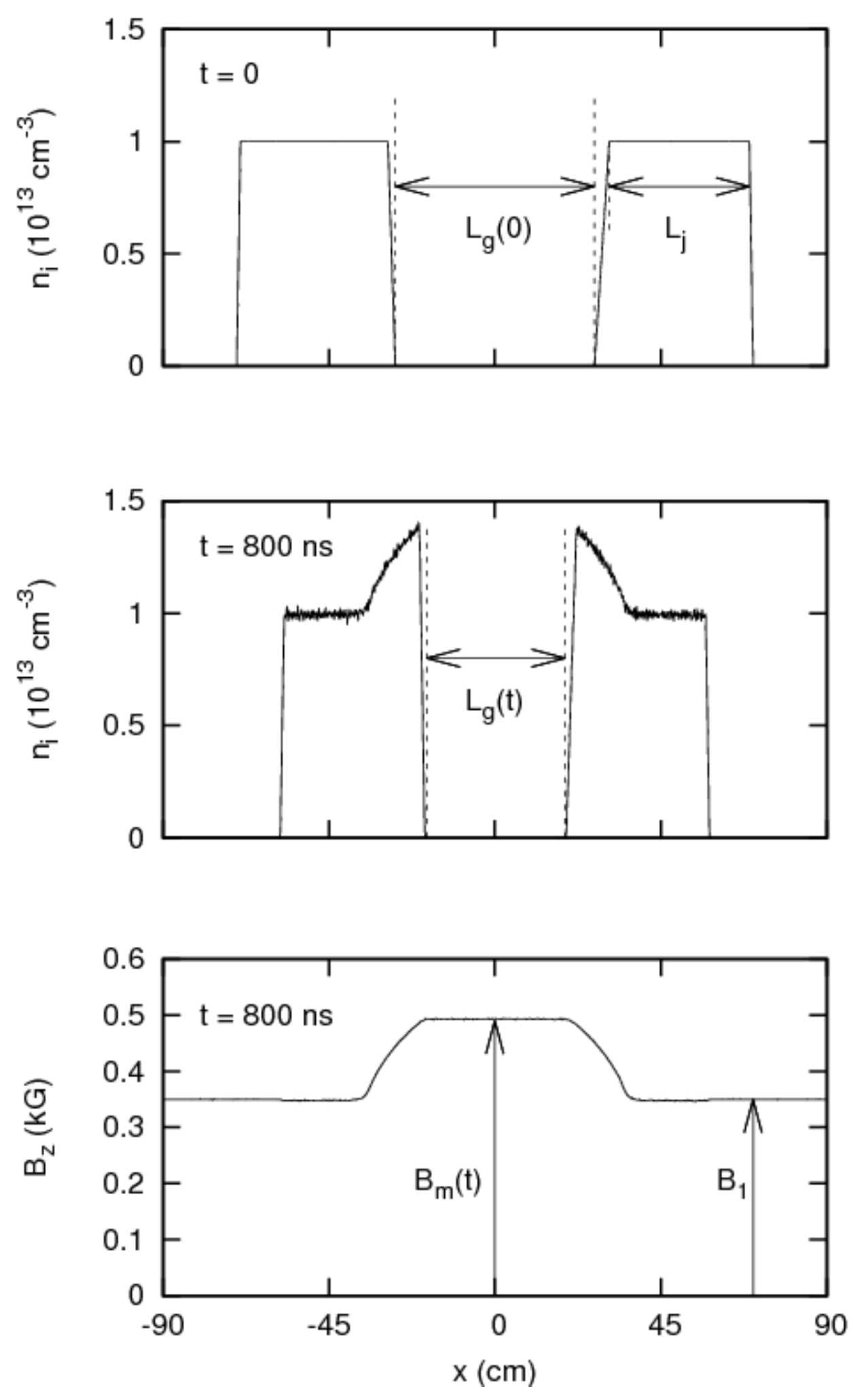}
\end{center}
\Caption[CT PC e:/reports/cless\_shock/jet\_geo.eps]{Two-jet
  simulation with an applied field of $350$~G and an initial jet
  separation of $\sim 45$~cm. The full initial conditions are given in
  Table~\ref{tab_merge}. The ion density is shown at (top) $t = 0$,
  and (middle) 800~ns. The magnetic field is also shown (bottom) at $t
  = 800$~ns.}
\label{fg:jet_geo}
\end{figure}

A simple model can be used to explain the early-time behavior
(pre-shock) of the 1D counter-propagating jet simulations with an
initial jet separation. To facilitate the discussion, we introduce
some nomenclature in Fig.~\ref{fg:jet_geo}. The initial gap separation at 
at $t = 0$ (top plot in figure) is $L_g(0)$, and the instantaneous gap separation $L_g(t)$ is
shown in the middle plot, which shows the ion density
after 800~ns. We also define the time-dependent magnetic field
strength at the midpoint between the jets as $B_m(t)$.

%%%%%%%%%%%%%%%%%%%%%%%%%%%%%%%%%%%%%%%%%%%%%%%%%%%%%%%%%%%%%%%%%%%%%%%%
%  Figure of 1D 2 Jet propagation with initial separation. Define Lg,
%  Bm , etc.

%
%  Figure taken from Sargas ~/hyperv/kinetic/weibel/run135
%  run135 is lsp run dir
%
%    gnuplot script ./jetgeoplot  (data files listed in there) 
%   eps file ./jet_geo.eps
%
%
%  Gnuplot figures were huge.
%  Open eps file with GV, copy into paint crop and save as png file 
%  Very inelegant but it was the best way I found to do it.
%  LAST STEP: POP won't take png files. Use convert to go to pdf
%
%%%%%%%%%%%%%%%%%%%%%%%%%%%%%%%%%%%%%%%%%%%%%%%%%%%%%%%%%%%%%%%%%%%%%%%%

\begin{figure}
\begin{center}
\includegraphics[clip = true,width=12cm]{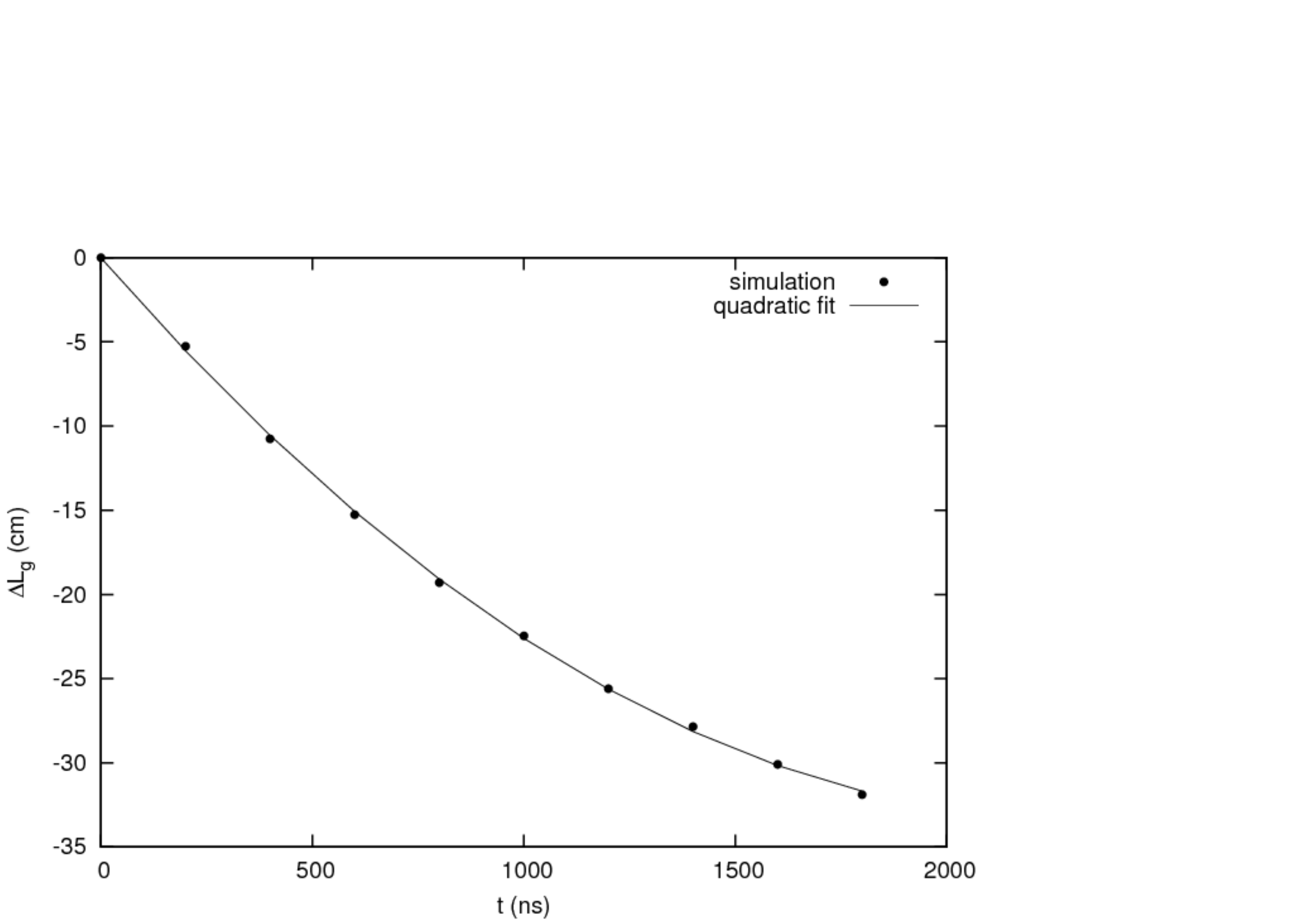}
\end{center}
\Caption[CT PC e:/reports/cless\_shock/lg.eps]{Change in jet
  separation distance $\Delta L_g$ as a function of time. The
  simulation data is compared to a quadratic fit based on
  Eq.~(\ref{eq:accel}).}
\label{fg:lg}
\end{figure}

In all these simulations (before the onset of shock waves), we observe
flux conservation, i.e.,
\begin{equation}
	L_g(t) \times B_m(t) \sim \mbox{constant},
\label{eq:flux}
\end{equation}
and a uniform acceleration of gap distance:
\begin{equation}
	L_g(t) \sim  L_g(0) -  2v_j t +  1/2 a t^2,
\label{eq:accel}
\end{equation}
where $a > 0$ is a constant acceleration value. This can be seen
explicitly in Fig.~\ref{fg:lg}, which plots $\Delta L_g [= L_g(t)
-L_g(0)]$ as a function of time for a simulation with $v_j =
150$~km/s, $L_g(0) = 94$~cm, $L_j = 42$~cm, and $B_1 = 350$~G\@. We have
also plotted $\Delta L_g$ as given by Eq.~(\ref{eq:accel}) with a
constant acceleration of $a = 12$~cm$/\mu$s$^2$, which was
determined by least-squares fitting.

%%%%%%%%%%%%%%%%%%%%%%%%%%%%%%%%%%%%%%%%%%%%%%%%%%%%%%%%%%%%%%%%%%%%%%%%
%  Figure of Delta L_g(t) compared with quadratic fit
%  Bm , etc.  a = ?
%
%  Figure taken from Sargas ~/hyperv/kinetic/weibel/run134 (lsp run directory)
%
%    gnuplot script ./plotstuff  (data files listed in there) 
%   eps file ./lg.eps
%
%
%  Gnuplot figures were huge.
%  Open eps file with GV, copy into paint crop and save as png file 
%  Very inelegant but it was the best way I found to do it.
%  LAST STEP: POP won't take png files. Use convert to go to pdf
%
%%%%%%%%%%%%%%%%%%%%%%%%%%%%%%%%%%%%%%%%%%%%%%%%%%%%%%%%%%%%%%%%%%%%%%%%

\begin{figure}
\begin{center}
\includegraphics[clip = true, width = 10 cm]{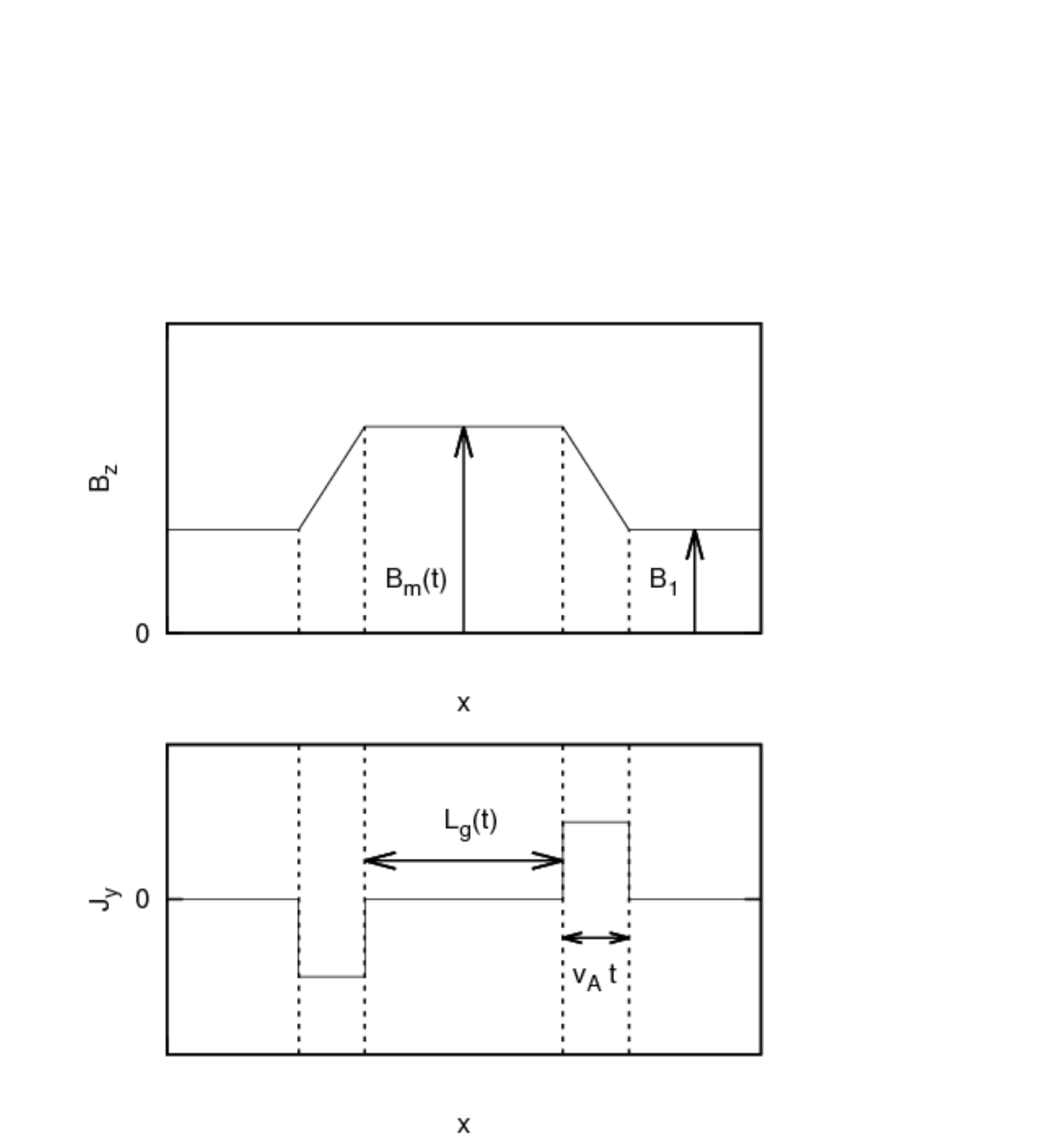}
\end{center}
\Caption[CT PC e:/reports/cless\_shock/jetscheme.eps]{Schematic view
of \Alf{} wave propagation in merging jet simulations.}
\label{fg:jetscheme}
\end{figure}

To derive an estimate for the gap acceleration, we consider a
simplified model of the 1D simulations. This is shown in
Fig.~\ref{fg:jetscheme}. We neglect displacement current, assume
linear ramps on $B_z$, and a piecewise constant current $J_y$.  By
imposing flux conservation and assuming a $\vec{J} \times \vec{B}$
force ($\beta \ll 1$) on the plasma in the transition region, we obtain
\begin{equation}
 a = 4 v_A v_j/L_g(0),
\label{eq:atheory}
\end{equation} 
where $v_A$ is calculated using the applied field $B_1$.

%%%%%%%%%%%%%%%%%%%%%%%%%%%%%%%%%%%%%%%%%%%%%%%%%%%%%%%%%%%%%%%%%%%%%%%%
%  Figure of 1D 2 Jet propagation with initial separation. Define Lg,
%  Bm , etc.
%
%  Figure taken from Sargas ~/hyperv/kinetic/weibel/
%
%    gnuplot script ./plotscheme  (data files listed in there) 
%   eps file ./jet_scheme.eps
%
%  Gnuplot figures were huge.
%  Open eps file with GV, copy into paint crop and save as png file 
%  Very inelegant but it was the best way I found to do it.
%  LAST STEP: POP won't take png files. Use convert to go to pdf
%
%%%%%%%%%%%%%%%%%%%%%%%%%%%%%%%%%%%%%%%%%%%%%%%%%%%%%%%%%%%%%%%%%%%%%%%%

In Table~\ref{tab_accel} we list the acceleration measured in a series
of simulations with varying values of $B_1$, $L_g(0)$, $L_j$, and
$v_j$. According to the simple theory of Eq.~(\ref{eq:atheory}) the
normalized acceleration $a L_g(0)/v_A v_j$ should be equal to 4.  The
\LSP{} simulations are seen to be in relatively good agreement with
the simple scaling derived above.
If we define the minimum jet separation as $L_{min}$, our
simple theory predicts that
\begin{equation}
  L_{min} /L_g(0)= 1 -  v_j/2v_A.
\end{equation}
This implies that the jets will not overlap unless $v_j > 2 v_A$. But,
as we see in Fig.~\ref{fg:n_xpx_2jet_1000sep.eps}, the shock waves
start before $L_g$ can reach 0.

%%%%%%%%%%%%%%%%%%%%%%%%%%%%%%%%%%%%%%%%%%%%%%%%%%%%%%%%%%%%%%%%%%%%%%%%
% Table of Alfenic acceleration
%
%
\begin{table}
\caption{Comparison of 1D simulation results to simple theory
  [Eq.~(\ref{eq:atheory})] of Alfv\'{e}n wave propagation into jets
  with a large initial separation.}
\begin{ruledtabular}
\begin{tabular}{ccccc}
$B_1 (G)$ & $L_g(0)$ (cm) & $L_j$ (cm) & $v_j$ (km/s) & $a L_g(0)/v_Av_j$ \\ \hline 
$1000$ & 94 & 42 & 150 & 2.90 \\
$350$ & 94 & 42 & 150 & 3.23 \\
$350$ & 54 & 42 & 150 & 2.79 \\
$350$ & 94 & 42 & 300 & 4.22 \\
$350$ & 54 & 62 & 150 & 3.0 \\
\end{tabular}
\end{ruledtabular}
\label{tab_accel}
\end{table}
%
%
%  Gnuplot figures were huge.
%  Open eps file with GV, copy into paint crop and save as png file 
%  Very inelegant but it was the best way I found to do it.
%  LAST STEP: POP won't take png files. Use convert to go to pdf
%
%%%%%%%%%%%%%%%%%%%%%%%%%%%%%%%%%%%%%%%%%%%%%%%%%%%%%%%%%%%%%%%%%%%%%%%%

\subsection{Unmagnetized and magnetized jet merging in 2D}
\label{sec:2d-simul-magn}

In this section we consider the results of 2D Cartesian simulations of
counter-propagating jets in perpendicular magnetic fields. The jet
propagation remains in the $x$ direction, and we simulate the 2D jets
in the $x$--$y$~plane. In 2D, we now have to explicitly specify the
direction of the perpendicular field in the $y$--$z$ plane. Details on
the simulation setup were given in Sec.~\ref{sec:setup-explicit-2d}.

\begin{figure}
\begin{center}
\includegraphics[clip = true, width = 10 cm]{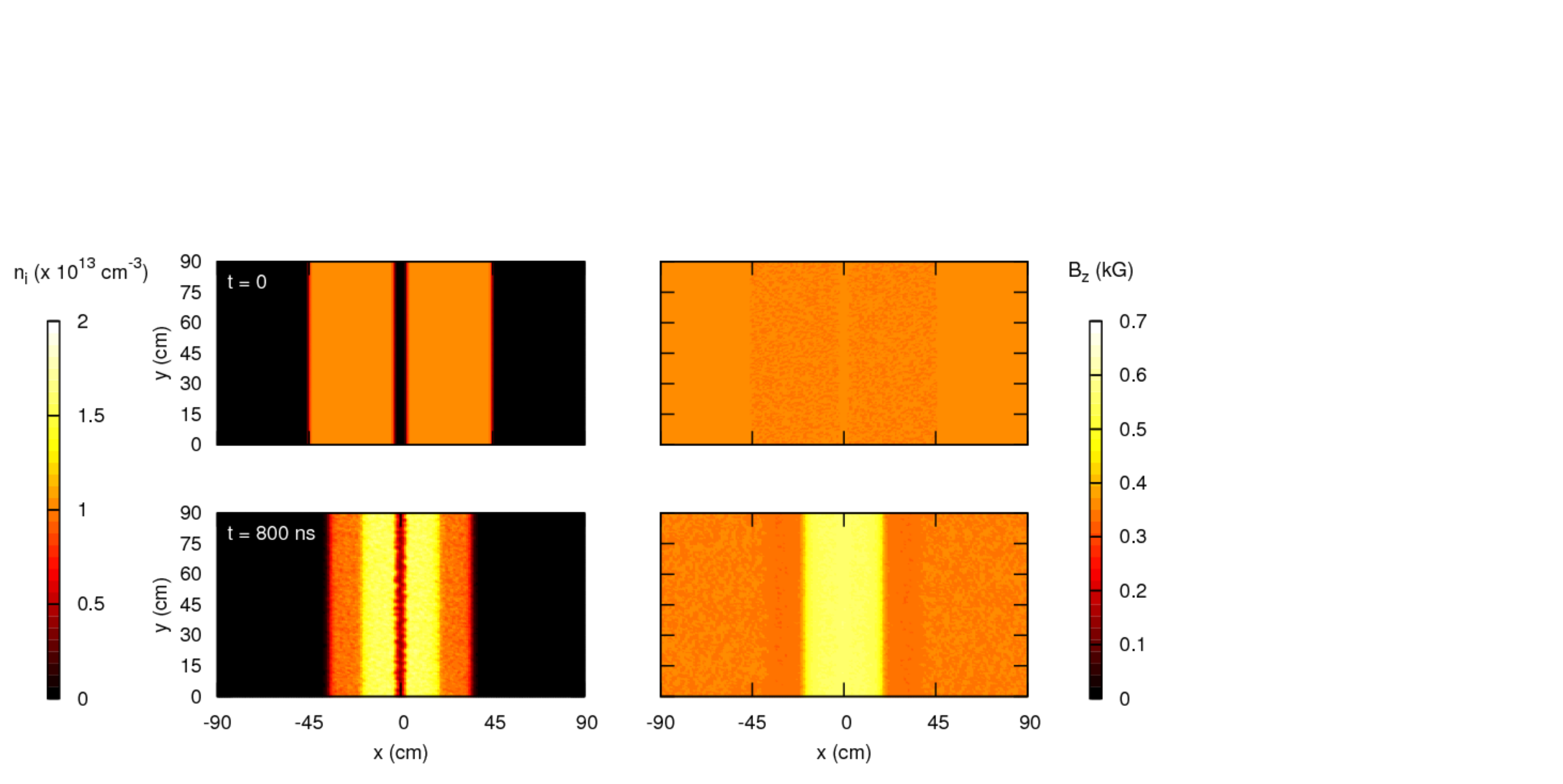}
\end{center}
\Caption[CT PC e:/reports/ni\_q2d.eps]{Results from 2D jet merging
  simulations. Initial jet parameters are given in
  Table~\ref{tab_merge}. The left column shows ion density contours
  and the right magnetic field strength $B_z$ at (top row) $t = 0$
  and (bottom row) $t = 800$~ns, well after shock formation.}
\label{fg:ni_q2d}
\end{figure}

We consider initially a quasi-1D simulation with no variation of any
physical quantity in the $y$ direction and periodic boundaries in
$y$. The initial density profiles can be seen in the top row of
Fig.~\ref{fg:ni_q2d}.  The full initial conditions for all the 2D
runs are given in Table~\ref{tab_merge}. The total $y$ extent is
many $d_i$ wide.  The initial conditions are $n_1
= 10^{13}$ cm$^{-3}$, $T_{e1} = T_{i1} = 1$~eV\@. The magnetic field is
in the $+z$ direction ( analogous to the 1D simulations as the
field is aligned in a virtual direction) with $B_1 = 350$~G and $v_j
= 150$ km/s.

We first note from the bottom row of Fig.~\ref{fg:ni_q2d}, which shows
$n_i$ and $B_z$ contours at $t = 800$~ns, well after the shock
formation, that there is no strongly evident structure in the $y$
direction in the bulk of the jets, although the $y$ extent is many $d_i$
wide. There is some small density variation along the $x =
0$ line where the density is very low, but these variations are on
the order of a cell size and are probably due to particle noise.

Due to the coarse spatial grid, we can no longer resolve the
oscillatory structure along the $x$-direction in the shock transition
region, which were seen in the highly resolved 1D simulations (e.g., see
Fig.~\ref{fg:n_xpx_2jet_1kg}).  But we obtain the same
shock speed (i.e., $M$) and density discontinuity,
$n_2/n_1 (= B_2/B_1 = v_1/v_2)$, as the corresponding 1D simulation.

%%%%%%%%%%%%%%%%%%%%%%%%%%%%%%%%%%%%%%%%%%%%%%%%%%%%%%%%%%%%%%%%%%%%%%%%
%  Figure of quasi 2D jet simulation
%
%  Figure taken from Sargas ~/hyperv/kinetic/weibel/run140 (lsp run directory)
%
%    gnuplot script ./splot_ff_bw  (data files listed in there) 
%   eps file ./ni_q2d.eps
%
%   Dump data from 2D contour plots in p4. Have to do some
%  post-processing to get gnuplot to read them (see notes in gnuplot
%  script)
%  run little shell script ~/bin/p4fix.p4 This is hard coded for 200 x
%  200 contour plot (Lsp default)
%
% n1 = 1013 cm-3
% (Te)1 = (Ti)1 = 2 eV
% B1 = 350
% vj = 150 km/s
%
%
%  Gnuplot figures were huge.
%  Open eps file with GV, copy into paint crop and save as png file 
%  Very inelegant but it was the best way I found to do it.
%  LAST STEP: POP won't take png files. Use convert to go to pdf
%
%%%%%%%%%%%%%%%%%%%%%%%%%%%%%%%%%%%%%%%%%%%%%%%%%%%%%%%%%%%%%%%%%%%%%%%%

\begin{figure}
\begin{center}
\includegraphics[clip = true, width = 10 cm]{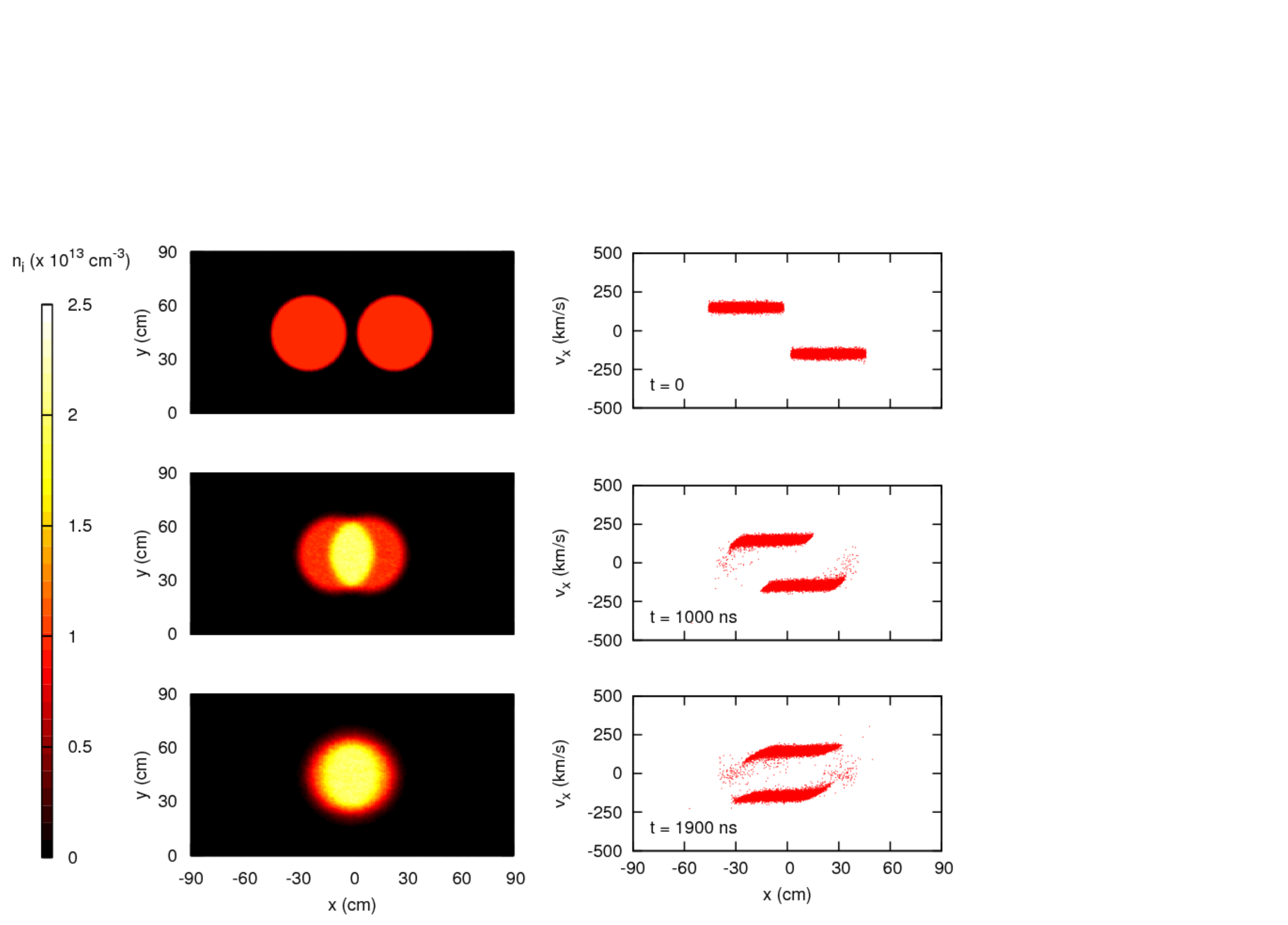}
\end{center}
\Caption[CT PC e:/reports/ni\_2dnob.eps]{Simulation (2D) of two
  counter-propagating plasma jets in the absence of a magnetic
  field. Ion density contours and ion $x$--$v_x$ phase-space data are
  shown at various times. There is essentially no interaction between
  the two jets.}
\label{fg:ni_2dnob}
\end{figure}

We now consider two uniform density disks with no applied magnetic
field. The periodic boundaries in $y$ have been retained. Ion density
contours and $x$--$v_x$ particle phase-space data are shown in
Fig.~\ref{fg:ni_2dnob}. The results are analogous to the unmagnetized
1D case, i.e., there is very little interaction between the jets in this
parameter regime in the absence of an applied magnetic field. There is
certainly no evidence of Weibel-induced unmagnetized shock waves on
time scales $\sim  1$~$\mu$s.
% Q would I have the resolution to see it? I'm resolving ion skin
% depths but not electron skin depths. 

% Figure poster slide 12
%%%%%%%%%%%%%%%%%%%%%%%%%%%%%%%%%%%%%%%%%%%%%%%%%%%%%%%%%%%%%%%%%%%%%%%%
%   2D two jet simulation with no field
%
%  Figure taken from Sargas ~/hyperv/kinetic/weibel/run185 (lsp run directory)
%
%    gnuplot script ./splot_ff_bw  (data files listed in there) 
%   eps file ./ni_2dnob.eps
%
%   Dump data from 2D contour plots in p4. Have to do some
%  post-processing to get gnuplot to read them (see notes in gnuplot
%  script)
%  run little shell script ~/bin/p4fix.p4 This is hard coded for 200 x
%  200 contour plot (Lsp default)
%
% n1 = 1013 cm-3
% (Te)1 = (Ti)1 = 2 eV
% B1 = 350
% vj = 150 km/s
%
%
%  Gnuplot figures were huge.
%  Open eps file with GV, copy into paint crop and save as png file 
%  Very inelegant but it was the best way I found to do it.
%  LAST STEP: POP won't take png files. Use convert to go to pdf
%
%%%%%%%%%%%%%%%%%%%%%%%%%%%%%%%%%%%%%%%%%%%%%%%%%%%%%%%%%%%%%%%%%%%%%%%%

\begin{figure}
\begin{center}
\includegraphics[clip = true, width = 10 cm]{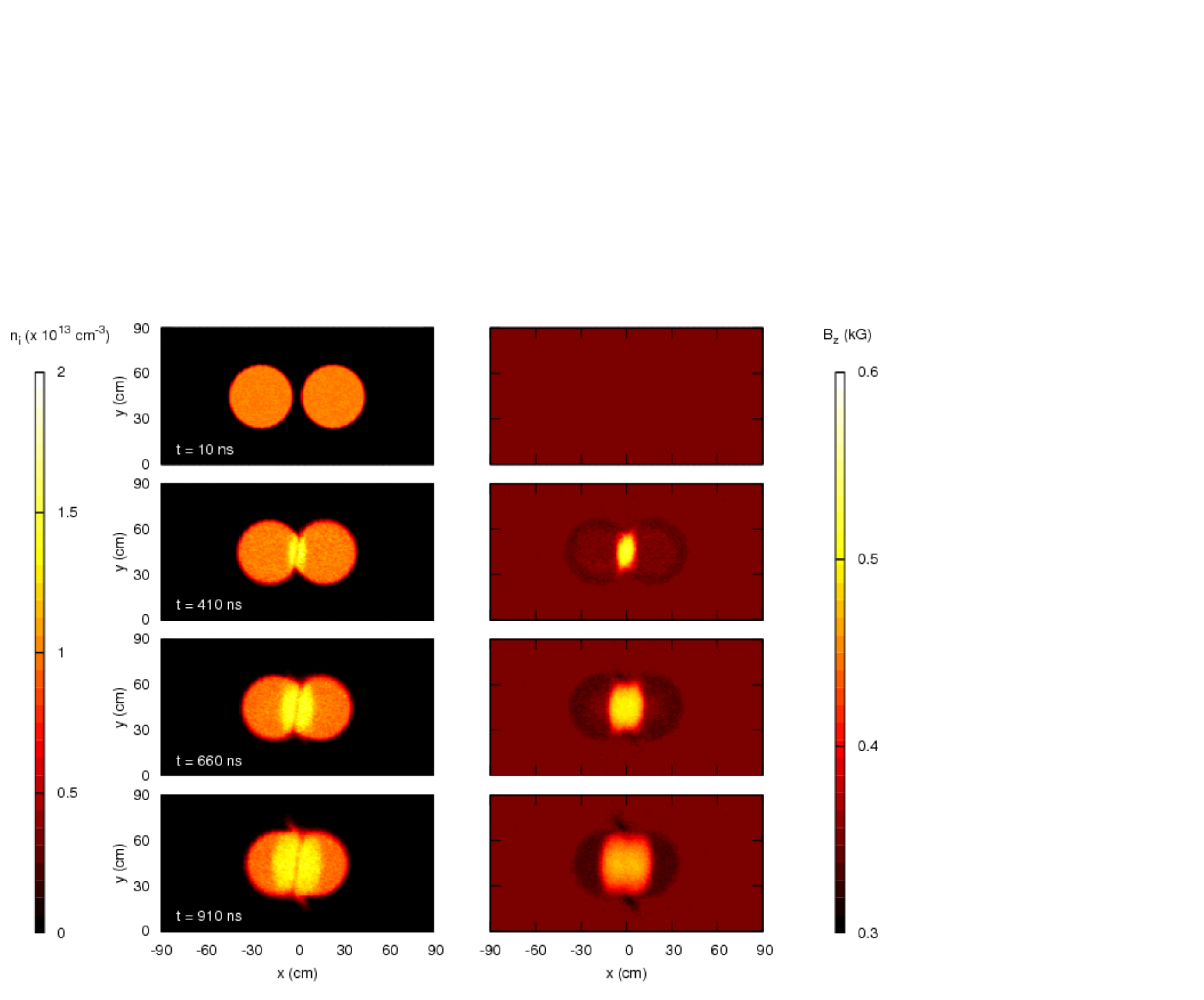}
\end{center}
\Caption[CT PC e:/reports/ni\_2dwb.eps]{Simulation (2D) of two
  counter-propagating plasma jets with a 350-G magnetic field in the
  $+z$ direction. Ion density and $B_z$ contours are shown at various
  times.}
\label{fg:ni_2dwb}
\end{figure}

We repeat the previous simulation but now add a 350-G magnetic field
in the virtual direction ($+z$). The $n_i$ and $B_z$ contours
are shown at various times in Fig.~\ref{fg:ni_2dwb}. Along the line $y
= 45$~cm, we find approximate values of $n_2/n_1 \sim B_2/B_1 \sim
1.5$, and $v_{1x}/v_{2x} \sim 2.4$, and a shock velocity of $v_s \sim
210$~km/s. Although the problem is now inherently 2D, we
nonetheless estimate $M \sim 1.4$ from the 1D
formula of Eq.~(\ref{eq:mach}). 

% Same something about why n2/n1 != v2/v1 it's a 2D problem now!
% Stats not great.  

We notice, however, that there is a conspicuous difference between
this simulation and the 1D and quasi-1D (Fig.~\ref{fg:ni_q2d})
analogs. Namely, the density at the center-of-mass of the two jets,
along the line at $x = 0$, remains large ($\geq
10^{13}$~cm$^{-3}$) throughout the simulation, indicating at least
some jet interpenetration at this point. In the 1D analogs, the density
remained close to zero at the jet center of mass.

\begin{figure}
\begin{center}
\includegraphics[clip = true, width = 10 cm]{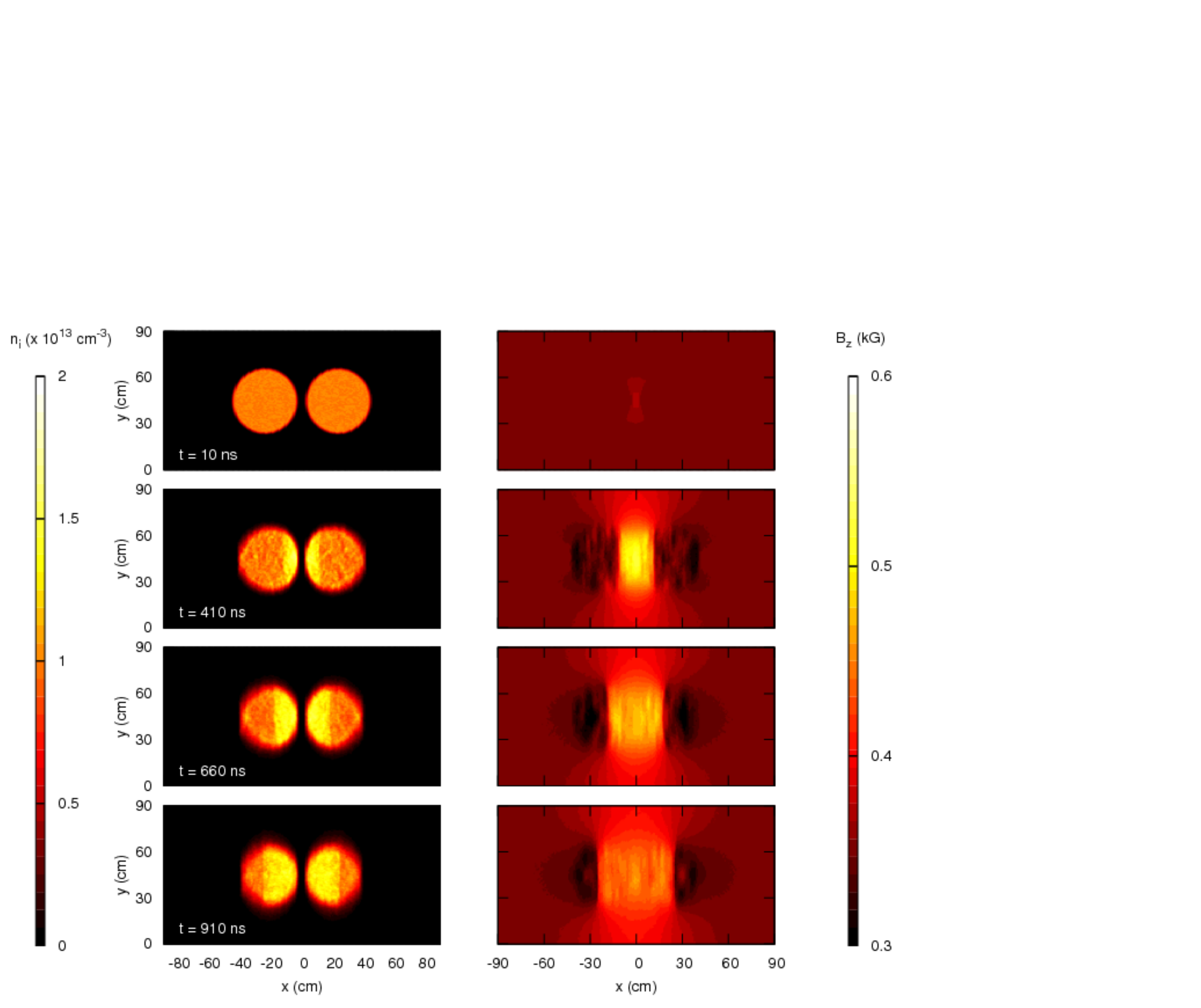}
\end{center}
\Caption[CT PC e:/reports/ni\_2dwby.eps]{Simulation (2D) of two
  counter-propagating plasma jets with a 350-G magnetic field in the
  $+y$ direction. Ion density and $B_y$ contours are shown at various
  times.}
\label{fg:ni_2dwby}
\end{figure}

In order to explain this qualitatively different behavior, we consider
first the results for a 2D simulation in which the 350-G field is
rotated from the virtual $+z$ to the $+y$ direction, which lies in the
simulation plane. These results are shown in Fig.~\ref{fg:ni_2dwby}.
For this simulation we find, along $y = 45$~cm, that $n_2/n_1 \sim
B_2/B_1 \sim v_1/v_2 \sim 1.5$ and $M \sim 1.6$. For this magnetic
field orientation, the results are qualitatively similar to the 1D
case, as $n_i$ remains zero along $x = 0$.

When the magnetic field is in the $y$ direction
(Fig.~\ref{fg:ni_2dwby}), the currents which drive the magnetic field,
$B_2 > B_1$, can flow in the virtual $z$ direction and remain
localized near $x = 0$~cm. This allows for large $\vec{J} \times
\vec{B}$ forces to reflect the incoming the jets. When the magnetic
field is in the virtual $z$-direction (Fig.~\ref{fg:ni_2dwb}) the
magnetic field is supported by finite circular current paths around
the perimeter of the bulk of the jet. The density gradient in $y$ acts
to minimize the $x$ component of the $\vec{J} \times \vec{B}$ force on
the jets.

%%%%%%%%%%%%%%%%%%%%%%%%%%%%%%%%%%%%%%%%%%%%%%%%%%%%%%%%%%%%%%%%%%%%%%%%
%   2D two jet simulation with Bz = 350
%
%  Figure taken from Sargas ~/hyperv/kinetic/weibel/run141 (lsp run directory)
%
%    gnuplot script ./splot_2dwb  (data files listed in there) 
%   eps file ./ni_2dnwb.eps
%
%   Dump data from 2D contour plots in p4. Have to do some
%  post-processing to get gnuplot to read them (see notes in gnuplot
%  script)
%  run little shell script ~/bin/p4fix.p4 This is hard coded for 200 x
%  200 contour plot (Lsp default)
%
%
%  Gnuplot figures were huge.
%  Open eps file with GV, copy into paint crop and save as png file 
%  Very inelegant but it was the best way I found to do it.
%  LAST STEP: POP won't take png files. Use convert to go to pdf
%
%
%  n1 = 1013 cm-3
%  (Te)1 = (Ti)1 = 2 eV
%  vj = 150 km/s
%  B1 (+z) = 350 G 
%
%%%%%%%%%%%%%%%%%%%%%%%%%%%%%%%%%%%%%%%%%%%%%%%%%%%%%%%%%%%%%%%%%%%%%%%%

%%%%%%%%%%%%%%%%%%%%%%%%%%%%%%%%%%%%%%%%%%%%%%%%%%%%%%%%%%%%%%%%%%%%%%%%
%   2D two jet simulation with By = 350
%
%  Figure taken from Sargas ~/hyperv/kinetic/weibel/run142 (lsp run directory)
%
%    gnuplot script ./splot_2dwby  (data files listed in there) 
%   eps file ./ni_2dnwby.eps
%
%   Dump data from 2D contour plots in p4. Have to do some
%  post-processing to get gnuplot to read them (see notes in gnuplot
%  script)
%  run little shell script ~/bin/p4fix.p4 This is hard coded for 200 x
%  200 contour plot (Lsp default)
%
%
%  Gnuplot figures were huge.
%  Open eps file with GV, copy into paint crop and save as png file 
%  Very inelegant but it was the best way I found to do it.
%  LAST STEP: POP won't take png files. Use convert to go to pdf
%
%  n1 = 1013 cm-3
%  (Te)1 = (Ti)1 = 2 eV
%  vj = 150 km/s
%  B1 (+y) = 350 G 
%
%%%%%%%%%%%%%%%%%%%%%%%%%%%%%%%%%%%%%%%%%%%%%%%%%%%%%%%%%%%%%%%%%%%%%%%%

\begin{figure}
\begin{center}
\includegraphics[clip = true, width = 10 cm]{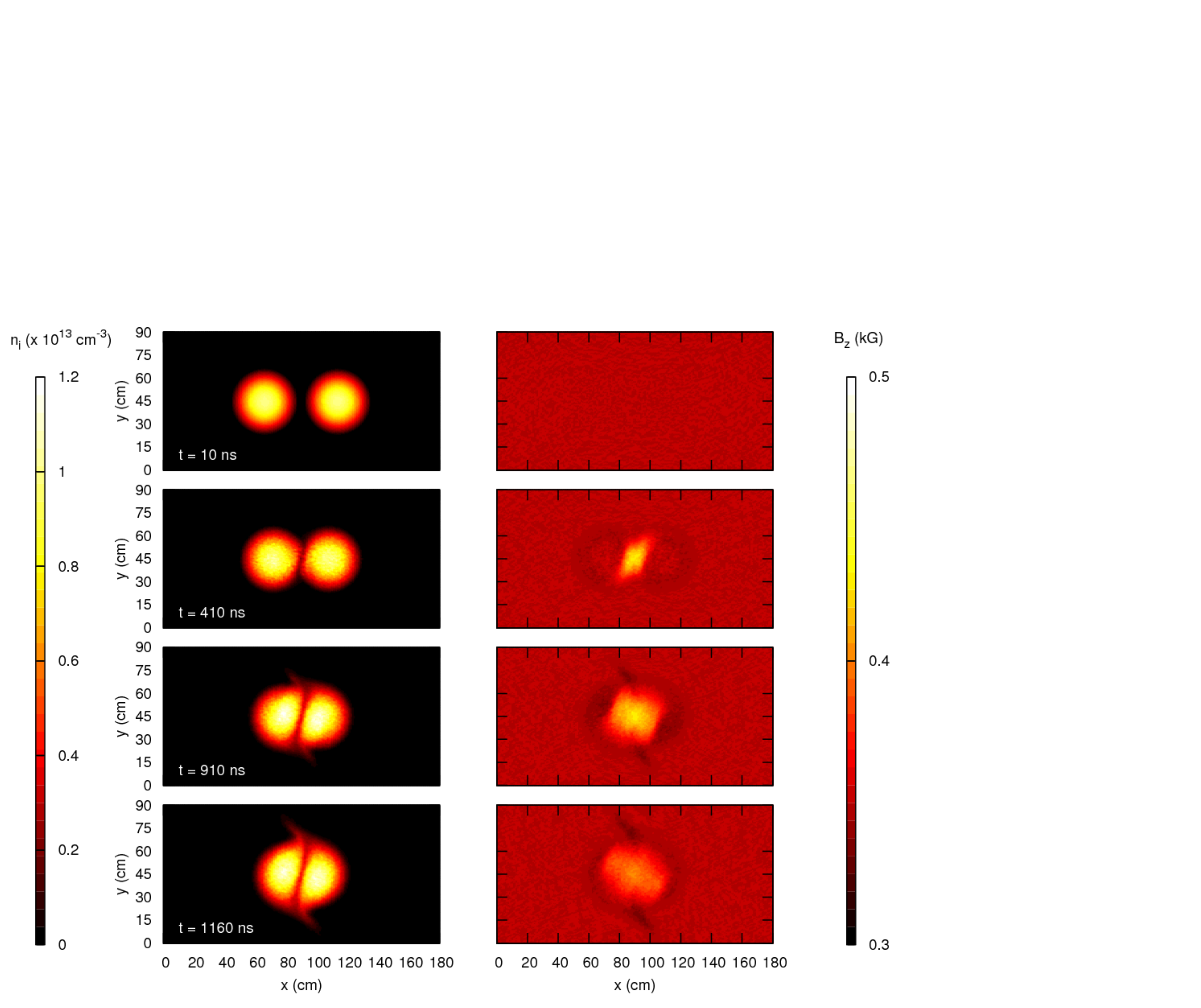}
\end{center}
\Caption[CT PC e:/reports/ni\_2dgrad.eps]{Simulation (2D) of two
  counter-propagating plasma jets with a 350-G magnetic field in the
  $+y$ direction. Ion density and $B_z$ contours are shown at various
  times. In contrast to Fig.~\ref{fg:ni_2dwb}, the jets have
  finite initial density gradients.}
\label{fg:ni_2dgrad}
\end{figure}

As a final simulation, we consider a 2D simulation with more realistic
density gradients. The 350-G field is in the $+z$ direction. The
initial $n_i$ contours can be seen in the top left of
Fig.~\ref{fg:ni_2dgrad}. From the $n_i$ and $B_z$ contours at
later times, we note that it is difficult to clearly see the
propagation front in the density contours as they are imposed on top
of the initial gradient, but the magnetic field front of the shock is
clearly seen in the field contours. The simulation results from
Figs.~\ref{fg:n_xpx_2jet_varyn} and \ref{fg:ni_2dwby} demonstrate that
shock structure can be observed in magnetized jets with realistic
density gradients.

%%%%%%%%%%%%%%%%%%%%%%%%%%%%%%%%%%%%%%%%%%%%%%%%%%%%%%%%%%%%%%%%%%%%%%%%
%   2D two jet simulation with Bz = 350 density gradients
%
%  Figure taken from Sargas ~/hyperv/kinetic/weibel/run143 (lsp run directory)
%
%    gnuplot script ./splot_2dgrad  (data files listed in there) 
%   eps file ./ni_2dgrad.eps
%
%   Dump data from 2D contour plots in p4. Have to do some
%  post-processing to get gnuplot to read them (see notes in gnuplot
%  script)
%  run little shell script ~/bin/p4fix.p4 This is hard coded for 200 x
%  200 contour plot (Lsp default)
%
%
%  Gnuplot figures were huge.
%  Open eps file with GV, copy into paint crop and save as png file 
%  Very inelegant but it was the best way I found to do it.
%  LAST STEP: POP won't take png files. Use convert to go to pdf
%
%%%%%%%%%%%%%%%%%%%%%%%%%%%%%%%%%%%%%%%%%%%%%%%%%%%%%%%%%%%%%%%%%%%%%%%%

\section{Conclusions}\label{sec:conclusions}

We have performed 1D and 2D PIC simulations of hydrogen plasma jet
propagation and head-on merging. In the parameter regime of the LANL
experiment, unmagnetized collisionless shocks could not be detected in
the simulation results.  The simulations do demonstrate the formation
of magnetized (perpendicular) collisionless shocks when $v_j < v_A$
($\beta \ll 1$). This requires that the jets be immersed in a field 
 $\sim 0.1$--1~kG\@. Simulations predict $M\approx 1$--2
 in this parameter range. Non-shock jet interactions (i.e., the \Alf{}
wave propagation discussed in Sec.~\ref{sec:effects-real-dens}) are
also observed in the simulations as well as ion kinetic effects.
These simulation confirm that the application of an
appropriate magnetic field to the LANL experiment is required.
Some simple calculations as well as preliminary
simulations show that an applied field cannot fully diffuse into the oncoming
jets on time scales $< 1$~$\mu$s. For this
reason the jets need to be ``born'' in the magnetic field or
penetrate into the field by some other mechanism than magnetic
diffusion~\cite{baker:65}. Since applied
fields were shown to suppress ambipolar diffusion [see
Fig.~\ref{fg:n_xpx}(c)], it would be valuable to perform a large-scale 2D
jet propagation simulation to see how jets would evolve when born or
injected into applied fields.  Larger-scale 2D or 3D simulations
require better spatial resolution to avoid numerical difficulties for
longer simulation times ($t > 1$~$\mu$s). Although not considered in
this paper, we point out that the LANL experiment should also be able to
observe unmagnetized {\em collisional} shocks with higher density
merging jets and has observed such phenomena
\cite{merritt:13}.

\begin{acknowledgments}
This work was supported by the Laboratory Directed Research and
Development (LDRD) Program at LANL through U.S. Dept.\ of Energy
contract no.~DE-AC52-06NA25396. The authors also acknowledge useful
discussions with Dr.\ D.~V.~Rose and Dr.\ N.~L.~Bennett of Voss
Scientific.
\end{acknowledgments}

% Create the reference section using BibTeX:
%\bibliography{strings,other,group}

\begin{thebibliography}{39}
\expandafter\ifx\csname natexlab\endcsname\relax\def\natexlab#1{#1}\fi
\expandafter\ifx\csname bibnamefont\endcsname\relax
  \def\bibnamefont#1{#1}\fi
\expandafter\ifx\csname bibfnamefont\endcsname\relax
  \def\bibfnamefont#1{#1}\fi
\expandafter\ifx\csname citenamefont\endcsname\relax
  \def\citenamefont#1{#1}\fi
\expandafter\ifx\csname url\endcsname\relax
  \def\url#1{\texttt{#1}}\fi
\expandafter\ifx\csname urlprefix\endcsname\relax\def\urlprefix{URL }\fi
\providecommand{\bibinfo}[2]{#2}
\providecommand{\eprint}[2][]{\url{#2}}

\bibitem[{\citenamefont{Sagdeev}(1966)}]{sagdeev66}
\bibinfo{author}{\bibfnamefont{R.~Z.} \bibnamefont{Sagdeev}},
  \bibinfo{journal}{Rev.\ Plasma Phys.} \textbf{\bibinfo{volume}{4}},
  \bibinfo{pages}{23} (\bibinfo{year}{1966}).

\bibitem[{\citenamefont{Sagdeev and Kennel}(1991)}]{sagdeev91}
\bibinfo{author}{\bibfnamefont{R.~Z.} \bibnamefont{Sagdeev}} \bibnamefont{and}
  \bibinfo{author}{\bibfnamefont{C.~F.} \bibnamefont{Kennel}},
  \bibinfo{journal}{Sci.\ Amer.} \textbf{\bibinfo{volume}{264}},
  \bibinfo{pages}{106} (\bibinfo{year}{1991}).

\bibitem[{\citenamefont{Ness et~al.}(1964)\citenamefont{Ness, Scearce, and
  Seek}}]{ness64}
\bibinfo{author}{\bibfnamefont{N.~F.} \bibnamefont{Ness}},
  \bibinfo{author}{\bibfnamefont{C.~S.} \bibnamefont{Scearce}},
  \bibnamefont{and} \bibinfo{author}{\bibfnamefont{J.~B.} \bibnamefont{Seek}},
  \bibinfo{journal}{J. Geophys.\ Res.} \textbf{\bibinfo{volume}{69}},
  \bibinfo{pages}{3531} (\bibinfo{year}{1964}).

\bibitem[{pla()}]{plasma2010}
\bibinfo{note}{S. C. Cowley and J. Peoples (co-chairs) and Plasma 2010
  Committee, {\em Plasma Science--Advancing Knowledge in the National Interest}
  (National Academies Press, Washington, D.C., 2007).}

\bibitem[{wop()}]{wopa10}
\bibinfo{note}{{\em Research Opportunities in Plasma Astrophysics}, Report of
  the Workshop on Opportunities in Plasma Astrophysics, Princeton, NJ,
  Jan.~18--21, 2010; download at http://www.pppl.gov/conferences/2010/WOPA.}

\bibitem[{\citenamefont{Podgornyi and Sagdeev}(1969)}]{podgornyi69}
\bibinfo{author}{\bibfnamefont{I.~M.} \bibnamefont{Podgornyi}}
  \bibnamefont{and} \bibinfo{author}{\bibfnamefont{R.~Z.}
  \bibnamefont{Sagdeev}}, \bibinfo{journal}{Sov.\ Phys.\ Uspekhi}
  \textbf{\bibinfo{volume}{98}}, \bibinfo{pages}{445} (\bibinfo{year}{1969}).

\bibitem[{\citenamefont{Zakharov}(2003)}]{zakharov03}
\bibinfo{author}{\bibfnamefont{Y.~P.} \bibnamefont{Zakharov}},
  \bibinfo{journal}{IEEE Trans.\ Plasma Sci.} \textbf{\bibinfo{volume}{31}},
  \bibinfo{pages}{1243} (\bibinfo{year}{2003}).

\bibitem[{\citenamefont{Woolsey et~al.}(2001)\citenamefont{Woolsey,
  \mbox{Y.~Abou~Ali}, Evans, \mbox{R. A. D. Grundy}, Pestehe, Carolan, Conway,
  Dendy, Helander, McClements et~al.}}]{woolsey01}
\bibinfo{author}{\bibfnamefont{N.~C.} \bibnamefont{Woolsey}},
  \bibinfo{author}{\bibnamefont{\mbox{Y.~Abou~Ali}}},
  \bibinfo{author}{\bibfnamefont{R.~G.} \bibnamefont{Evans}},
  \bibinfo{author}{\bibnamefont{\mbox{R. A. D. Grundy}}},
  \bibinfo{author}{\bibfnamefont{S.~J.} \bibnamefont{Pestehe}},
  \bibinfo{author}{\bibfnamefont{P.~G.} \bibnamefont{Carolan}},
  \bibinfo{author}{\bibfnamefont{N.~J.} \bibnamefont{Conway}},
  \bibinfo{author}{\bibfnamefont{R.~O.} \bibnamefont{Dendy}},
  \bibinfo{author}{\bibfnamefont{P.}~\bibnamefont{Helander}},
  \bibinfo{author}{\bibfnamefont{K.~G.} \bibnamefont{McClements}},
  \bibnamefont{et~al.}, \bibinfo{journal}{Phys.\ Plasmas}
  \textbf{\bibinfo{volume}{8}}, \bibinfo{pages}{2439} (\bibinfo{year}{2001}).

\bibitem[{\citenamefont{Horton et~al.}(2007)\citenamefont{Horton, Chiu,
  Ditmire, Valanju, Presura, Ivanov, Sentoku, Sotnikov, Esaulov, \mbox{N. Le
  Galloudec} et~al.}}]{horton07}
\bibinfo{author}{\bibfnamefont{W.}~\bibnamefont{Horton}},
  \bibinfo{author}{\bibfnamefont{C.}~\bibnamefont{Chiu}},
  \bibinfo{author}{\bibfnamefont{T.}~\bibnamefont{Ditmire}},
  \bibinfo{author}{\bibfnamefont{P.}~\bibnamefont{Valanju}},
  \bibinfo{author}{\bibfnamefont{R.}~\bibnamefont{Presura}},
  \bibinfo{author}{\bibfnamefont{V.~V.} \bibnamefont{Ivanov}},
  \bibinfo{author}{\bibfnamefont{Y.}~\bibnamefont{Sentoku}},
  \bibinfo{author}{\bibfnamefont{V.~I.} \bibnamefont{Sotnikov}},
  \bibinfo{author}{\bibfnamefont{A.}~\bibnamefont{Esaulov}},
  \bibinfo{author}{\bibnamefont{\mbox{N. Le Galloudec}}}, \bibnamefont{et~al.},
  \bibinfo{journal}{Advances Space Res.} \textbf{\bibinfo{volume}{39}},
  \bibinfo{pages}{358} (\bibinfo{year}{2007}).

\bibitem[{\citenamefont{Ponomarenko et~al.}(2008)\citenamefont{Ponomarenko,
  \mbox{Yu.~P.~Zakharov}, Antonov, Boyarintsev, Melekhov, Posukh,
  Shaikhislamov, and Vchivkov}}]{ponomarenko08}
\bibinfo{author}{\bibfnamefont{A.~G.} \bibnamefont{Ponomarenko}},
  \bibinfo{author}{\bibnamefont{\mbox{Yu.~P.~Zakharov}}},
  \bibinfo{author}{\bibfnamefont{V.~M.} \bibnamefont{Antonov}},
  \bibinfo{author}{\bibfnamefont{E.~L.} \bibnamefont{Boyarintsev}},
  \bibinfo{author}{\bibfnamefont{A.~V.} \bibnamefont{Melekhov}},
  \bibinfo{author}{\bibfnamefont{V.~G.} \bibnamefont{Posukh}},
  \bibinfo{author}{\bibfnamefont{I.~F.} \bibnamefont{Shaikhislamov}},
  \bibnamefont{and} \bibinfo{author}{\bibfnamefont{K.~V.}
  \bibnamefont{Vchivkov}}, \bibinfo{journal}{Plasma Phys.\ Control.\ Fus.}
  \textbf{\bibinfo{volume}{50}}, \bibinfo{pages}{074015}
  (\bibinfo{year}{2008}).

\bibitem[{\citenamefont{Romagnani et~al.}(2008)\citenamefont{Romagnani,
  Bulanov, Borghesi, Audebert, Gauthier, Lowenbruck, Mackinnon, Patel,
  Pretzler, Toncian et~al.}}]{romagnani08}
\bibinfo{author}{\bibfnamefont{L.}~\bibnamefont{Romagnani}},
  \bibinfo{author}{\bibfnamefont{S.~V.} \bibnamefont{Bulanov}},
  \bibinfo{author}{\bibfnamefont{M.}~\bibnamefont{Borghesi}},
  \bibinfo{author}{\bibfnamefont{P.}~\bibnamefont{Audebert}},
  \bibinfo{author}{\bibfnamefont{J.~C.} \bibnamefont{Gauthier}},
  \bibinfo{author}{\bibfnamefont{K.}~\bibnamefont{Lowenbruck}},
  \bibinfo{author}{\bibfnamefont{A.~J.} \bibnamefont{Mackinnon}},
  \bibinfo{author}{\bibfnamefont{P.}~\bibnamefont{Patel}},
  \bibinfo{author}{\bibfnamefont{G.}~\bibnamefont{Pretzler}},
  \bibinfo{author}{\bibfnamefont{T.}~\bibnamefont{Toncian}},
  \bibnamefont{et~al.}, \bibinfo{journal}{Phys.\ Rev.\ Lett.}
  \textbf{\bibinfo{volume}{101}}, \bibinfo{pages}{025004}
  (\bibinfo{year}{2008}).

\bibitem[{\citenamefont{Constantin et~al.}(2009)\citenamefont{Constantin,
  Gekelman, Pribyl, Everson, Schaeffer, Kugland, Presura, Neff, Plechaty,
  Vincena et~al.}}]{constantin09}
\bibinfo{author}{\bibfnamefont{C.}~\bibnamefont{Constantin}},
  \bibinfo{author}{\bibfnamefont{W.}~\bibnamefont{Gekelman}},
  \bibinfo{author}{\bibfnamefont{P.}~\bibnamefont{Pribyl}},
  \bibinfo{author}{\bibfnamefont{E.}~\bibnamefont{Everson}},
  \bibinfo{author}{\bibfnamefont{D.}~\bibnamefont{Schaeffer}},
  \bibinfo{author}{\bibfnamefont{N.}~\bibnamefont{Kugland}},
  \bibinfo{author}{\bibfnamefont{R.}~\bibnamefont{Presura}},
  \bibinfo{author}{\bibfnamefont{S.}~\bibnamefont{Neff}},
  \bibinfo{author}{\bibfnamefont{C.}~\bibnamefont{Plechaty}},
  \bibinfo{author}{\bibfnamefont{S.}~\bibnamefont{Vincena}},
  \bibnamefont{et~al.}, \bibinfo{journal}{Astrophys.\ Space Sci.}
  \textbf{\bibinfo{volume}{322}}, \bibinfo{pages}{155} (\bibinfo{year}{2009}).

\bibitem[{\citenamefont{Kuramitsu et~al.}(2011)\citenamefont{Kuramitsu, Sakawa,
  Morita, Gregory, Waugh, Dono, Aoki, Tanji, Koenig, Woolsey
  et~al.}}]{kuramitsu11}
\bibinfo{author}{\bibfnamefont{Y.}~\bibnamefont{Kuramitsu}},
  \bibinfo{author}{\bibfnamefont{Y.}~\bibnamefont{Sakawa}},
  \bibinfo{author}{\bibfnamefont{T.}~\bibnamefont{Morita}},
  \bibinfo{author}{\bibfnamefont{C.~D.} \bibnamefont{Gregory}},
  \bibinfo{author}{\bibfnamefont{J.~N.} \bibnamefont{Waugh}},
  \bibinfo{author}{\bibfnamefont{S.}~\bibnamefont{Dono}},
  \bibinfo{author}{\bibfnamefont{H.}~\bibnamefont{Aoki}},
  \bibinfo{author}{\bibfnamefont{H.}~\bibnamefont{Tanji}},
  \bibinfo{author}{\bibfnamefont{M.}~\bibnamefont{Koenig}},
  \bibinfo{author}{\bibfnamefont{N.}~\bibnamefont{Woolsey}},
  \bibnamefont{et~al.}, \bibinfo{journal}{Phys.\ Rev.\ Lett.}
  \textbf{\bibinfo{volume}{106}}, \bibinfo{pages}{175002}
  (\bibinfo{year}{2011}).

\bibitem[{\citenamefont{Ross et~al.}(2012)\citenamefont{Ross, Glenzer, Amendt,
  Berger, Divol, Kugland, Landen, Plechaty, Remington, Ryutov et~al.}}]{ross12}
\bibinfo{author}{\bibfnamefont{J.~S.} \bibnamefont{Ross}},
  \bibinfo{author}{\bibfnamefont{S.~H.} \bibnamefont{Glenzer}},
  \bibinfo{author}{\bibfnamefont{P.}~\bibnamefont{Amendt}},
  \bibinfo{author}{\bibfnamefont{R.}~\bibnamefont{Berger}},
  \bibinfo{author}{\bibfnamefont{L.}~\bibnamefont{Divol}},
  \bibinfo{author}{\bibfnamefont{N.~L.} \bibnamefont{Kugland}},
  \bibinfo{author}{\bibfnamefont{O.~L.} \bibnamefont{Landen}},
  \bibinfo{author}{\bibfnamefont{C.}~\bibnamefont{Plechaty}},
  \bibinfo{author}{\bibfnamefont{B.}~\bibnamefont{Remington}},
  \bibinfo{author}{\bibfnamefont{D.}~\bibnamefont{Ryutov}},
  \bibnamefont{et~al.}, \bibinfo{journal}{Phys.\ Plasmas}
  \textbf{\bibinfo{volume}{19}}, \bibinfo{pages}{056501}
  (\bibinfo{year}{2012}).

\bibitem[{\citenamefont{Schaeffer et~al.}(2012)\citenamefont{Schaeffer,
  Everson, Winske, Constantin, Bondarenko, Morton, Flippo, Montgomery,
  Gaillard, and Niemann}}]{schaeffer12}
\bibinfo{author}{\bibfnamefont{D.~B.} \bibnamefont{Schaeffer}},
  \bibinfo{author}{\bibfnamefont{E.~T.} \bibnamefont{Everson}},
  \bibinfo{author}{\bibfnamefont{D.}~\bibnamefont{Winske}},
  \bibinfo{author}{\bibfnamefont{C.~G.} \bibnamefont{Constantin}},
  \bibinfo{author}{\bibfnamefont{A.~S.} \bibnamefont{Bondarenko}},
  \bibinfo{author}{\bibfnamefont{L.~A.} \bibnamefont{Morton}},
  \bibinfo{author}{\bibfnamefont{K.~A.} \bibnamefont{Flippo}},
  \bibinfo{author}{\bibfnamefont{D.~S.} \bibnamefont{Montgomery}},
  \bibinfo{author}{\bibfnamefont{S.~A.} \bibnamefont{Gaillard}},
  \bibnamefont{and} \bibinfo{author}{\bibfnamefont{C.}~\bibnamefont{Niemann}},
  \bibinfo{journal}{Phys.\ Plasmas} \textbf{\bibinfo{volume}{19}},
  \bibinfo{pages}{070702} (\bibinfo{year}{2012}).

\bibitem[{\citenamefont{Swadling et~al.}(2013)\citenamefont{Swadling, Lebedev,
  Niasse, Chittenden, Hall, Suzuki-Vidal, Burdiak, Harvey-Thompson, Bland,
  \mbox{P. De Grouchy} et~al.}}]{swadling13}
\bibinfo{author}{\bibfnamefont{G.~F.} \bibnamefont{Swadling}},
  \bibinfo{author}{\bibfnamefont{S.~V.} \bibnamefont{Lebedev}},
  \bibinfo{author}{\bibfnamefont{N.}~\bibnamefont{Niasse}},
  \bibinfo{author}{\bibfnamefont{J.~P.} \bibnamefont{Chittenden}},
  \bibinfo{author}{\bibfnamefont{G.~N.} \bibnamefont{Hall}},
  \bibinfo{author}{\bibfnamefont{F.}~\bibnamefont{Suzuki-Vidal}},
  \bibinfo{author}{\bibfnamefont{G.}~\bibnamefont{Burdiak}},
  \bibinfo{author}{\bibfnamefont{A.~J.} \bibnamefont{Harvey-Thompson}},
  \bibinfo{author}{\bibfnamefont{S.~N.} \bibnamefont{Bland}},
  \bibinfo{author}{\bibnamefont{\mbox{P. De Grouchy}}}, \bibnamefont{et~al.},
  \bibinfo{journal}{Phys.\ Plasmas} \textbf{\bibinfo{volume}{20}},
  \bibinfo{pages}{022705} (\bibinfo{year}{2013}).

\bibitem[{\citenamefont{Drake}(2000)}]{drake00}
\bibinfo{author}{\bibfnamefont{R.~P.} \bibnamefont{Drake}},
  \bibinfo{journal}{Phys.\ Plasmas} \textbf{\bibinfo{volume}{7}},
  \bibinfo{pages}{4690} (\bibinfo{year}{2000}).

\bibitem[{\citenamefont{Moser et~al.}(2012)\citenamefont{Moser, Hsu, Dunn,
  Martens, Adams, Merritt, Lynn, Gilmore, Thoma, and Welch}}]{moser12}
\bibinfo{author}{\bibfnamefont{A.~L.} \bibnamefont{Moser}},
  \bibinfo{author}{\bibfnamefont{S.~C.} \bibnamefont{Hsu}},
  \bibinfo{author}{\bibfnamefont{J.~P.} \bibnamefont{Dunn}},
  \bibinfo{author}{\bibfnamefont{D.~T.} \bibnamefont{Martens}},
  \bibinfo{author}{\bibfnamefont{C.~S.} \bibnamefont{Adams}},
  \bibinfo{author}{\bibfnamefont{E.~C.} \bibnamefont{Merritt}},
  \bibinfo{author}{\bibfnamefont{A.~G.} \bibnamefont{Lynn}},
  \bibinfo{author}{\bibfnamefont{M.~A.} \bibnamefont{Gilmore}},
  \bibinfo{author}{\bibfnamefont{C.}~\bibnamefont{Thoma}}, \bibnamefont{and}
  \bibinfo{author}{\bibfnamefont{D.~R.} \bibnamefont{Welch}},
  \bibinfo{journal}{Bull.\ Amer.\ Phys.\ Soc.} \textbf{\bibinfo{volume}{57}},
  \bibinfo{pages}{130} (\bibinfo{year}{2012}).

\bibitem[{lsp()}]{lspcode}
\bibinfo{note}{{\LSP{}} was developed by ATK Mission Research Corporation with
  initial support from the Department of Energy (DOE) SBIR Program.}

\bibitem[{\citenamefont{Hughes et~al.}(1999)\citenamefont{Hughes, Clark, and
  Yu}}]{hughes:99c}
\bibinfo{author}{\bibfnamefont{T.~P.} \bibnamefont{Hughes}},
  \bibinfo{author}{\bibfnamefont{R.~E.} \bibnamefont{Clark}}, \bibnamefont{and}
  \bibinfo{author}{\bibfnamefont{S.~S.} \bibnamefont{Yu}},
  \bibinfo{journal}{Phys. Rev. ST Accel. Beams} \textbf{\bibinfo{volume}{2}},
  \bibinfo{pages}{110401} (\bibinfo{year}{1999}).

\bibitem[{\citenamefont{Shimada and Hoshino}(2000)}]{shimada:00}
\bibinfo{author}{\bibfnamefont{N.}~\bibnamefont{Shimada}} \bibnamefont{and}
  \bibinfo{author}{\bibfnamefont{M.}~\bibnamefont{Hoshino}},
  \bibinfo{journal}{"The Astrophysical Journal"}
  \textbf{\bibinfo{volume}{543}}, \bibinfo{pages}{L67} (\bibinfo{year}{2000}).

\bibitem[{\citenamefont{Tidman and Krall}(1971)}]{tidman:71}
\bibinfo{author}{\bibfnamefont{D.~A.} \bibnamefont{Tidman}} \bibnamefont{and}
  \bibinfo{author}{\bibfnamefont{N.~A.} \bibnamefont{Krall}},
  \emph{\bibinfo{title}{Shock Waves in Collisionless Plasmas}}
  (\bibinfo{publisher}{John Wiley and Sons}, \bibinfo{address}{New York},
  \bibinfo{year}{1971}).

\bibitem[{\citenamefont{Weibel}(1959)}]{weibel59}
\bibinfo{author}{\bibfnamefont{E.~S.} \bibnamefont{Weibel}},
  \bibinfo{journal}{Phys.\ Rev.\ Lett.} \textbf{\bibinfo{volume}{2}},
  \bibinfo{pages}{83} (\bibinfo{year}{1959}).

\bibitem[{\citenamefont{Kato and Takabe}(2008)}]{kato:08}
\bibinfo{author}{\bibfnamefont{T.~N.} \bibnamefont{Kato}} \bibnamefont{and}
  \bibinfo{author}{\bibfnamefont{H.}~\bibnamefont{Takabe}},
  \bibinfo{journal}{"The Astrophysical Journal"}
  \textbf{\bibinfo{volume}{681}}, \bibinfo{pages}{L93} (\bibinfo{year}{2008}).

\bibitem[{\citenamefont{Witherspoon et~al.}(2011)\citenamefont{Witherspoon,
  Brockington, Case, Messer, Wu, Elton, Hsu, Cassibry, Gilmore, and the
  PLX~Team}}]{witherspoon11}
\bibinfo{author}{\bibfnamefont{F.~D.} \bibnamefont{Witherspoon}},
  \bibinfo{author}{\bibfnamefont{S.}~\bibnamefont{Brockington}},
  \bibinfo{author}{\bibfnamefont{A.}~\bibnamefont{Case}},
  \bibinfo{author}{\bibfnamefont{S.~J.} \bibnamefont{Messer}},
  \bibinfo{author}{\bibfnamefont{L.}~\bibnamefont{Wu}},
  \bibinfo{author}{\bibfnamefont{R.}~\bibnamefont{Elton}},
  \bibinfo{author}{\bibfnamefont{S.~C.} \bibnamefont{Hsu}},
  \bibinfo{author}{\bibfnamefont{J.~T.} \bibnamefont{Cassibry}},
  \bibinfo{author}{\bibfnamefont{M.~A.} \bibnamefont{Gilmore}},
  \bibnamefont{and} \bibinfo{author}{\bibnamefont{the PLX~Team}},
  \bibinfo{journal}{Bull.\ Amer.\ Phys.\ Soc.} \textbf{\bibinfo{volume}{56}},
  \bibinfo{pages}{311} (\bibinfo{year}{2011}).

\bibitem[{\citenamefont{Thoma et~al.}(2011)\citenamefont{Thoma, Welch, Clark,
  Bruner, Mac{F}arlane, and Golovkin}}]{thoma:11}
\bibinfo{author}{\bibfnamefont{C.}~\bibnamefont{Thoma}},
  \bibinfo{author}{\bibfnamefont{D.~R.} \bibnamefont{Welch}},
  \bibinfo{author}{\bibfnamefont{R.~E.} \bibnamefont{Clark}},
  \bibinfo{author}{\bibfnamefont{N.}~\bibnamefont{Bruner}},
  \bibinfo{author}{\bibfnamefont{J.}~\bibnamefont{Mac{F}arlane}},
  \bibnamefont{and} \bibinfo{author}{\bibfnamefont{I.}~\bibnamefont{Golovkin}},
  \bibinfo{journal}{Phys. Plasmas} \textbf{\bibinfo{volume}{18}},
  \bibinfo{pages}{103507} (\bibinfo{year}{2011}).

\bibitem[{\citenamefont{Hsu et~al.}(2012{\natexlab{a}})\citenamefont{Hsu,
  Merritt, Moser, Awe, Brockington, Davis, Adams, Case, Cassibry, Dunn
  et~al.}}]{hsu12pop}
\bibinfo{author}{\bibfnamefont{S.~C.} \bibnamefont{Hsu}},
  \bibinfo{author}{\bibfnamefont{E.~C.} \bibnamefont{Merritt}},
  \bibinfo{author}{\bibfnamefont{A.~L.} \bibnamefont{Moser}},
  \bibinfo{author}{\bibfnamefont{T.~J.} \bibnamefont{Awe}},
  \bibinfo{author}{\bibfnamefont{S.~J.~E.} \bibnamefont{Brockington}},
  \bibinfo{author}{\bibfnamefont{J.~S.} \bibnamefont{Davis}},
  \bibinfo{author}{\bibfnamefont{C.~S.} \bibnamefont{Adams}},
  \bibinfo{author}{\bibfnamefont{A.}~\bibnamefont{Case}},
  \bibinfo{author}{\bibfnamefont{J.~T.} \bibnamefont{Cassibry}},
  \bibinfo{author}{\bibfnamefont{J.~P.} \bibnamefont{Dunn}},
  \bibnamefont{et~al.}, \bibinfo{journal}{Phys.\ Plasmas}
  \textbf{\bibinfo{volume}{19}}, \bibinfo{pages}{123514}
  (\bibinfo{year}{2012}{\natexlab{a}}).

\bibitem[{\citenamefont{Hsu et~al.}(2012{\natexlab{b}})\citenamefont{Hsu, Awe,
  Brockington, Case, Cassibry, Kagan, Messer, Stanic, Tang, Welch
  et~al.}}]{hsu12ieee}
\bibinfo{author}{\bibfnamefont{S.~C.} \bibnamefont{Hsu}},
  \bibinfo{author}{\bibfnamefont{T.~J.} \bibnamefont{Awe}},
  \bibinfo{author}{\bibfnamefont{S.}~\bibnamefont{Brockington}},
  \bibinfo{author}{\bibfnamefont{A.}~\bibnamefont{Case}},
  \bibinfo{author}{\bibfnamefont{J.~T.} \bibnamefont{Cassibry}},
  \bibinfo{author}{\bibfnamefont{G.}~\bibnamefont{Kagan}},
  \bibinfo{author}{\bibfnamefont{S.~J.} \bibnamefont{Messer}},
  \bibinfo{author}{\bibfnamefont{M.}~\bibnamefont{Stanic}},
  \bibinfo{author}{\bibfnamefont{X.}~\bibnamefont{Tang}},
  \bibinfo{author}{\bibfnamefont{D.~R.} \bibnamefont{Welch}},
  \bibnamefont{et~al.}, \bibinfo{journal}{IEEE Trans.\ Plasma Sci.}
  \textbf{\bibinfo{volume}{40}}, \bibinfo{pages}{1287}
  (\bibinfo{year}{2012}{\natexlab{b}}).

\bibitem[{\citenamefont{Genoni et~al.}(2004)\citenamefont{Genoni, Rose, Welch,
  and Lee}}]{genoni:04a}
\bibinfo{author}{\bibfnamefont{T.~C.} \bibnamefont{Genoni}},
  \bibinfo{author}{\bibfnamefont{D.~V.} \bibnamefont{Rose}},
  \bibinfo{author}{\bibfnamefont{D.~R.} \bibnamefont{Welch}}, \bibnamefont{and}
  \bibinfo{author}{\bibfnamefont{E.~P.} \bibnamefont{Lee}},
  \bibinfo{journal}{Phys. Plasmas} \textbf{\bibinfo{volume}{11}},
  \bibinfo{pages}{73} (\bibinfo{year}{2004}).

\bibitem[{\citenamefont{Rose et~al.}(2007)\citenamefont{Rose, Genoni, Welch,
  Startsev, and Davidson}}]{rose:07a}
\bibinfo{author}{\bibfnamefont{D.~V.} \bibnamefont{Rose}},
  \bibinfo{author}{\bibfnamefont{T.~C.} \bibnamefont{Genoni}},
  \bibinfo{author}{\bibfnamefont{D.~R.} \bibnamefont{Welch}},
  \bibinfo{author}{\bibfnamefont{E.~A.} \bibnamefont{Startsev}},
  \bibnamefont{and} \bibinfo{author}{\bibfnamefont{R.~C.}
  \bibnamefont{Davidson}}, \bibinfo{journal}{Phys. Rev. ST Accel. Beams}
  \textbf{\bibinfo{volume}{10}}, \bibinfo{pages}{034203}
  (\bibinfo{year}{2007}).

\bibitem[{\citenamefont{Rose et~al.}(2005)\citenamefont{Rose, Guillory, and
  Beall}}]{rose:05a}
\bibinfo{author}{\bibfnamefont{D.~V.} \bibnamefont{Rose}},
  \bibinfo{author}{\bibfnamefont{J.}~\bibnamefont{Guillory}}, \bibnamefont{and}
  \bibinfo{author}{\bibfnamefont{J.~H.} \bibnamefont{Beall}},
  \bibinfo{journal}{Phys. Plasmas} \textbf{\bibinfo{volume}{12}},
  \bibinfo{pages}{014501} (\bibinfo{year}{2005}).

\bibitem[{\citenamefont{Welch et~al.}(2011)\citenamefont{Welch, Rose, Thoma,
  Genoni, Bruner, Clark, Stygar, and Leeper}}]{welch:11}
\bibinfo{author}{\bibfnamefont{D.~R.} \bibnamefont{Welch}},
  \bibinfo{author}{\bibfnamefont{D.~V.} \bibnamefont{Rose}},
  \bibinfo{author}{\bibfnamefont{C.}~\bibnamefont{Thoma}},
  \bibinfo{author}{\bibfnamefont{T.}~\bibnamefont{Genoni}},
  \bibinfo{author}{\bibfnamefont{N.}~\bibnamefont{Bruner}},
  \bibinfo{author}{\bibfnamefont{R.~E.} \bibnamefont{Clark}},
  \bibinfo{author}{\bibfnamefont{W.}~\bibnamefont{Stygar}}, \bibnamefont{and}
  \bibinfo{author}{\bibfnamefont{R.}~\bibnamefont{Leeper}},
  \bibinfo{journal}{Bull. Am. Phys. Soc.} \textbf{\bibinfo{volume}{56}},
  \bibinfo{pages}{200} (\bibinfo{year}{2011}).

\bibitem[{\citenamefont{MacFarlane et~al.}(2007)\citenamefont{MacFarlane,
  Golovkin, Wang, Woodruff, and Pereyra}}]{macfarlane:07}
\bibinfo{author}{\bibfnamefont{J.~J.} \bibnamefont{MacFarlane}},
  \bibinfo{author}{\bibfnamefont{I.~E.} \bibnamefont{Golovkin}},
  \bibinfo{author}{\bibfnamefont{P.}~\bibnamefont{Wang}},
  \bibinfo{author}{\bibfnamefont{P.~R.} \bibnamefont{Woodruff}},
  \bibnamefont{and} \bibinfo{author}{\bibfnamefont{N.~A.}
  \bibnamefont{Pereyra}}, \bibinfo{journal}{High Energy Density Physics}
  \textbf{\bibinfo{volume}{3}}, \bibinfo{pages}{181} (\bibinfo{year}{2007}).

\bibitem[{\citenamefont{Zheng et~al.}(2000)\citenamefont{Zheng, Chen, and
  Zhang}}]{zheng:00}
\bibinfo{author}{\bibfnamefont{F.}~\bibnamefont{Zheng}},
  \bibinfo{author}{\bibfnamefont{Z.}~\bibnamefont{Chen}}, \bibnamefont{and}
  \bibinfo{author}{\bibfnamefont{J.}~\bibnamefont{Zhang}},
  \bibinfo{journal}{IEEE Trans. on Microwave Theory Tech.}
  \textbf{\bibinfo{volume}{48}}, \bibinfo{pages}{1550} (\bibinfo{year}{2000}).

\bibitem[{\citenamefont{Zheng and Chen}(2001)}]{zheng:01}
\bibinfo{author}{\bibfnamefont{F.}~\bibnamefont{Zheng}} \bibnamefont{and}
  \bibinfo{author}{\bibfnamefont{Z.}~\bibnamefont{Chen}},
  \bibinfo{journal}{IEEE Trans. on Microwave Theory Tech.}
  \textbf{\bibinfo{volume}{49}}, \bibinfo{pages}{1006} (\bibinfo{year}{2001}).

\bibitem[{\citenamefont{Birdsall and Langdon}(1991)}]{birdsall:91d}
\bibinfo{author}{\bibfnamefont{C.~K.} \bibnamefont{Birdsall}} \bibnamefont{and}
  \bibinfo{author}{\bibfnamefont{A.~B.} \bibnamefont{Langdon}},
  \emph{\bibinfo{title}{Plasma Physics Via Computer Simulation}}
  (\bibinfo{publisher}{Adam Hilger}, \bibinfo{address}{New York},
  \bibinfo{year}{1991}).

\bibitem[{\citenamefont{Rambo and Procassini}(1995)}]{rambo:95}
\bibinfo{author}{\bibfnamefont{P.~W.} \bibnamefont{Rambo}} \bibnamefont{and}
  \bibinfo{author}{\bibfnamefont{R.~J.} \bibnamefont{Procassini}},
  \bibinfo{journal}{Phys. Plasmas} \textbf{\bibinfo{volume}{2}},
  \bibinfo{pages}{3130} (\bibinfo{year}{1995}).

\bibitem[{\citenamefont{Baker and Hammel}(1965)}]{baker:65}
\bibinfo{author}{\bibfnamefont{D.}~\bibnamefont{Baker}} \bibnamefont{and}
  \bibinfo{author}{\bibfnamefont{J.}~\bibnamefont{Hammel}},
  \bibinfo{journal}{Phys. Fluids} \textbf{\bibinfo{volume}{8}},
  \bibinfo{pages}{713} (\bibinfo{year}{1965}).

\bibitem[{\citenamefont{Merritt et~al.}()\citenamefont{Merritt, Moser, Hsu,
  Loverich, and Gilmore}}]{merritt:13}
\bibinfo{author}{\bibfnamefont{E.~C.} \bibnamefont{Merritt}},
  \bibinfo{author}{\bibfnamefont{A.~L.} \bibnamefont{Moser}},
  \bibinfo{author}{\bibfnamefont{S.~C.} \bibnamefont{Hsu}},
  \bibinfo{author}{\bibfnamefont{J.}~\bibnamefont{Loverich}}, \bibnamefont{and}
  \bibinfo{author}{\bibfnamefont{M.~A.} \bibnamefont{Gilmore}},
  \bibinfo{note}{``Experimental characterization of the stagnation layer
  between two obliquely merging supersonic plasma jets,'' submitted for
  publication (2013)}.

\end{thebibliography}

% Manually inserted contents of bbl file (for submisssion)

\end{document}